\documentclass{aa}
\usepackage{graphicx}
\usepackage{blindtext}
\usepackage{txfonts}
\usepackage[colorlinks=true,linkcolor=blue,citecolor=blue,urlcolor=blue]{hyperref}
\usepackage{comment}
\usepackage{dcolumn}
\usepackage{soul}
\usepackage{arydshln} 
\usepackage{natbib}  
\newcommand{\degree}{\ensuremath{^\circ}}

\begin{document} 
   \title{Magnetic connectivity from the Sun to the Earth with MHD models}

   \subtitle{I. Impact of the magnetic modelling for connectivity validation}

   \author{S. Kennis
          \inst{1}
          \and
          B. Perri  \inst{1,2}
          \and 
          S. Poedts \inst{1,3}
          }

   \institute{Centre for Mathematical Plasma-Astrophysics, KU Leuven, Celestijnenlaan 200B,
3001 Leuven, Belgium\\
         \and
         AIM/DAp - CEA Paris-Saclay, Université Paris-Saclay, Université Paris-Cité, Gif-sur-Yvette, France
        \and 
        Institute of Physics, University of Maria Curie-Skłodowska, Pl.\ Marii Curie-Skłodowska 5, 20-031 Lublin, Poland
            }

   \date{} 

 
  \abstract  
   {The magnetic connectivity between the Sun and the Earth is crucial to understanding solar wind and space weather events. However, it is challenging to determine because of the lack of direct observations, which explains the need for reliable simulations. }
   {The most used method in the last few years is the two-step ballistic method, but it has a lot of free parameters that can affect the final result. Thus, we want to provide a connectivity method based on self-consistent magnetohydrodynamics (MHD) models.}
   {To do so, we combine the COCONUT coronal model with the EUHFORIA heliospheric model to compute the magnetic field lines from the Earth to the Sun. We develop a way to quantify the uncertainty associated with this computation, both spatial and temporal. To validate our method, we selected 4 cases already studied in the literature and associated with High-Speed-Stream events  coming from unambiguous coronal holes visible on the disk.}
   {We always find a partial overlap with the assumed CH of origin: 19\% for event 1, 100\% for event 2, 45\% for event 3 and 100\% for event 4. We looked at the polarity at Earth over the full Carrington rotation to better understand these results. We found that, on average, MHD simulations provide a very good polarity estimation: 69\% agreement with real data for event 1, 36\% for event 2, 68\% for event 3 and 69\% for event 4. For events 1 and 3, we can then explain the mixed results by the spatial and temporal uncertainty. An interesting result is that, for MHD models, minimum activity cases appear to be more challenging because of multiple recurrent crossings of the HCS, while maximum activity cases appear easier because of the latitudinal extent of the HCS. A similar result was also found in \citet{Badman2023} with Parker Solar Probe data.
   }
   {We thus demonstrate that it is possible to use MHD models to compute magnetic connectivity and that it provides results just as good as the two-step ballistic method, with additional possibilities for improvements as the models integrate more critical physics.}

   \keywords{Corona, Magnetic fields, Connectivity, Solar wind, MHD}

   \maketitle
%

\section{Introduction}

Space weather can be described as the ability to anticipate the most energetic events from the Sun and their corresponding impact on our planet \citep{Schrijver2015_cospar}. Such events include, for example, solar particles accelerated to a relativistic speed by flares (Solar Energetic Particles or SEPs) or large clouds of coronal matter ejected in the interplanetary medium by magnetic reconnection (Coronal Mass Ejections or CMEs) \citep{Temmer2021}. These events can have a strong impact on the Earth's environment by disrupting the magnetosphere's magnetic field and enhancing the radiations in the van Allen belts \citep{Pulkkinen2007}. These consequences are significant for our technological society, as they can jeopardise astronaut equipment, satellite lifetimes, communication efficiency, aviation crew safety and even ground-based large-scale electrical installations \citep{Lanzerotti2001}. However, connecting these solar events with their geo-effective counterpart is far from trivial \citep{Zhang2021}: this is due to the dynamics of the interplanetary medium, shaped by the continuous ejection of particles that is the solar wind \citep{Parker1958}. The latter can slow down or even deflect events \citep{Lavraud2014}, and even cause disturbances by itself through its fastest component (High-Speed Streams or HSSs) \citep{Verbanac2011}. When the solar wind becomes highly dynamic (for example, close to maximum activity), it then becomes especially challenging to produce reliable space weather forecasts \citep{Riley2018}. 

One tool to connect solar observations with the ones at L1 is to use the interplanetary magnetic field \citep{Owens2013}. It is generated inside the Sun through the dynamo effect, i.e.\ the ability of a magnetised fluid to sustain and amplify a large-scale magnetic field against Ohmic dissipation \citep{Moffatt1978, Parker1993}. It is variable in time, with an 11-year cycle of activity for sunspots and a 22-year cycle for polarity reversal \citep{Hathaway2015}. This large-scale magnetic field expands across the solar surface and bathes the entire solar system. 
 It then interacts with the solar wind, creating complex dynamics where the particles follow the field lines in the lower corona. In contrast, further away from the Sun, the frozen magnetic field traces the path of the plasma. Energetic particles emitted during specific events such as SEPs are also sensitive to the influence of magnetic field and tend to gyrate around field lines \citep{Heber2006}.
Knowing the interplanetary magnetic field can thus give a map for the Sun-Earth connectivity. A direct application of this effect is the Parker spiral. Due to the Sun's rotation, magnetic field lines are twisted to an angle that reaches 45 degrees at the Earth \citep{Parker1958}, and hence western limb events tend to be more geo-effective on average \citep{Cid2012}. However, this method is still challenging to use, as there are no direct observations of the interplanetary magnetic field (only punctual local in situ measurements or indirect proxies such as white-light polarised brightness images) \citep{Owens2013}. Most methods thus rely on models, but the latter usually do not consider the perturbations created by turbulence, waves, shears, shocks or magnetic reconnection.

One of the first examples of the computation of the magnetic connectivity can be found in \cite{Nolte1973} with the ERQH (Extrapolated Quasi-Radial Hypervelocity) approximation. This method is based on the Parker spiral with a constant solar wind speed along the trajectory and was used to trace back to the solar surface the origin of an HSS \citep{Krieger1973} and a SEP events \citep{Roelof1973}. This method was later labelled the ballistic back-mapping method, as we start from the planet/spacecraft and compute the solar wind trajectory. \cite{Neugebauer1998} was the following study to go beyond to compute the connectivity of WIND and Ulysses spacecraft, exploring new semi-empirical models such as the Potential Field Source Surface extrapolation (PFSS, see \cite{Schatten1969}) or current sheet models \citep{Schatten1971}. These new methods were labelled magnetic mappings, starting from the solar magnetic field and focusing on the extension of the field lines. More recent studies using Parker Solar Probe (PSP) data have shown that these methods alone have limitations \citep{MacNeil2022} and that the best way to compute the magnetic connectivity is to use both methods one after the other, what is now called the two-step ballistic mapping \citep{Peleikis2017}. Such methods are now used to compute the connectivity for solar missions such as Solar Orbiter \citep{Rouillard2020}. There is ongoing research on how to improve both the two-step ballistic method and the computation of its corresponding uncertainties \citep{koukras2022, DaSilva2023, Dakeyo2024}.  

However, these empirical models tend to have a lot of free parameters that will strongly impact the connectivity forecast \citep{Badman2020}. MHD models are more self-consistent but are usually not used for space-weather applications as they tend to be too slow for operational use. This assumption has changed recently with new models such as EUHFORIA \citep{Pomoell2018}, COCONUT \citep{Perri2022} and ICARUS \citep{Verbeke2022}, specifically designed and optimised for operational forecasts. This forecasting time can even be improved through the architecture of forecasting facilities, which allows the coupling of coronal and heliospheric models \citep{Poedts2020}. Through this chain of models, we can also derive the magnetic connectivity and then test the advantages and limitations of this new method compared to existing ones.
Such a comparison has been made in \cite{Neugebauer1998} with the MAS (Magnetohydrodynamic Algorithm outside a Sphere) from \cite{Linker1997}. 
 This work was completed by \citet{Riley2006}, which found that the topology for PFSS and MHD models are usually similar. More recently, \citet{Badman2023} studied the connectivity from the Sun to Parker Solar Probe using many different models, including MHD models such as MAS but also MS-FLUKSS, \citep{Singh2022}. They also found similar results for Encounter 4 and 10 for connectivity estimations made with PFSS or MHD models. However, such a systematic comparison has not been conducted recently for the connectivity to Earth.

The structure of this paper is as follows. In Section~\ref{sec:models}, we provide more information about the 3D MHD models used (COCONUT for the corona and EUHFORIA  for the heliosphere), how we connect the two to compute the  magnetic connectivity and the alternative way of computing the connectivity using PFSS and ballistic back-mapping that we used for comparison. In Section~\ref{sec:Caseselection}, we present the different validation cases selected. In Section~\ref{sec:Connectivity}, we compute the magnetic connectivity for the chosen cases and compare the results with traditional connectivity methods. In Section~\ref{sec:Polarity}, we discuss the quality of these results and explain their differences using in situ data. Finally, in Section~\ref{sec:Conclusion}, we summarise all results and offer our future perspectives on this study. 

\section{Computation of the magnetic connectivity}
\label{sec:models}
As explained above, this study is unique in that it uses only MHD models to compute the magnetic connectivity. 
\subsection{MHD models}
\label{sec:mhd_models}

\subsubsection{MHD coronal model: COCONUT}
\label{sec:COCONUT}
COCONUT (COolfluid COroNa UnsTructured) is a 3D MHD data-driven coronal model that uses implicit numerical methods. The full description of the model can be found in \citet{Perri2022}, as we recall here only its main characteristics. The model is based on the COOLFluid framework for fluid mechanics \citep{Lani2013}. It solves the ideal MHD equations written in their conservative form in Cartesian coordinates and dimensionless form.  It uses a time implicit backward Euler scheme for finite volume methods to reach operational running times \citep{Perri2022} and runs on an unstructured grid to avoid polar singularities \citep{Brchnelova2022}. The default mesh used a 6th-level subdivision of the geodesic polyhedron with 20,480 surface elements, resulting in a grid with 3.9M elements. To ensure the divergence constraint $\nabla \cdot \boldsymbol{B} = 0$, we use the Artificial Compressibility Analogy \citep{chorin1997}, which is very similar to the Hyperbolic Divergence Cleaning (HDC) method initially developed by \cite{Dedner2002}.
The model neglected some elements to speed up the calculations, like the Hall term in the induction equation. Also, the Coriolis force or centrifugal forces are neglected in the momentum equation. Hence, the solar rotation is not considered since it does not influence the result below 20 solar radii. \citep{Perri2022}. 
COCONUT takes as input a synoptic magnetic map of the Sun as the inner boundary condition for the radial magnetic field. The map is projected onto spherical harmonics and reconstructed using the first 30 modes. 
It has been validated both at minimum \citep{Perri2023} and at maximum of activity \citep{kuzma2023} by comparison with various solar data. 
 More specifically, the validation was done during total solar eclipse events, using white light polarised brightness images. That way, the coronal magnetic field structure could be inferred and compared with the numerically obtained magnetic field.

This version of the COCONUT model uses a polytropic relation between the density and the pressure as an approximation of coronal heating. Indeed, by stating that the corona is almost isothermal (with a polytropic index of 1.05), we simulate a constant input of energy to accelerate the wind. However, with this assumption, only a slow or fast solar wind is considered, and not both simultaneously. The model was run with the same parameters described in \cite{Perri2023}, which means $\rho_\odot = 1.67 \times 10^{-16}\ \mathrm{g/cm^3}$ and $T_\odot = 1.9 \times 10^6\ \mathrm{K}$ for fixed-value Dirichlet conditions of density and pressure. The pressure at the inner boundary follows from the solar surface temperature by application of the ideal gas law: $P_\odot = 4.15 \times 10^{-2} \, \mathrm{dyn/cm^2}$. All other quantities  of the magnetic field have zero-gradient conditions.  This means that $\partial B_{\theta} / \partial r = \partial B_\varphi /  \partial r = 0$. As for the velocity, it is aligned with the magnetic field in order to limit surface currents, as explained in \citet{Brchnelova2022b}. The frame used in COCONUT is the same as the input map. As we will explain later, we used SDO/HMI synoptic maps in the Carrington frame for this study. We have indeed demonstrated in \cite{Perri2023} that these maps yielded the most realistic results for the base parameters of COCONUT. 

\subsubsection{MHD heliospheric model: EUHFORIA}
\label{sec:EUHFORIA}
EUHFORIA (EUropean Heliospheric FORecasting Information Asset) is a model to forecast the solar wind in the inner heliosphere and potential CME evolution within \citep{Pomoell2018}. The complete model contains a coronal (up to 21.5 solar radii) and a heliospheric (up to Mars' orbit) part. However, we replaced the coronal part with COCONUT to provide the solar wind radial speed, temperature, density and radial magnetic field at the interface.  
The heliospheric part is a 3D magneto-hydrodynamics model in which the ideal MHD equations are solved, gravity included. These equations are solved in the Heliocentric Earth Equatorial coordinate system (HEEQ-coordinates). We use a finite volume method combined with a constrained transport approach to solve the MHD equations. We use an approximate Riemann solver with standard piece-wise linear reconstruction \citep{Kissmann2012, Pomoell2012} to obtain a robust and second-order accurate scheme. At the outer radial boundary, we use open boundary conditions implemented via a simple extrapolation, whereas at the latitudinal boundaries, we use symmetric reflection boundary conditions.
Although the chosen frame is not inertial, we choose to omit the Coriolis and centrifugal terms, which should be the result of Earth's orbital motion, as their contribution is negligible  compared to the other plasma forces and timescales due to the slow rotation of the Sun. A value of 1.5 is selected for the polytropic index as in \cite{Odstrcil2004}.
The default set-up has a 2$\degree$ angular resolution.
We did not include any CMEs in this project. 

\subsection{Magnetic Connectivity using MHD models}
\label{sec:connection}
This section now explains how we compute the magnetic connectivity from the Sun to the Earth with the outputs of the two previously described MHD models. As explained previously, both models have different frames (Carrington for COCONUT, HEEQ for EUHFORIA) that we must adjust to make the physical quantities continuous. The equivalent of the HEEQ frame for the coronal part is the Stonyhurst heliographic coordinate frame \citep{Thompson2006,John2010}. We thus compute the longitude of the Carrington longitude in this frame and proceed to rotate the COCONUT data accordingly. 

\begin{figure*}[h!]
    \centering
    \includegraphics[width=\hsize]{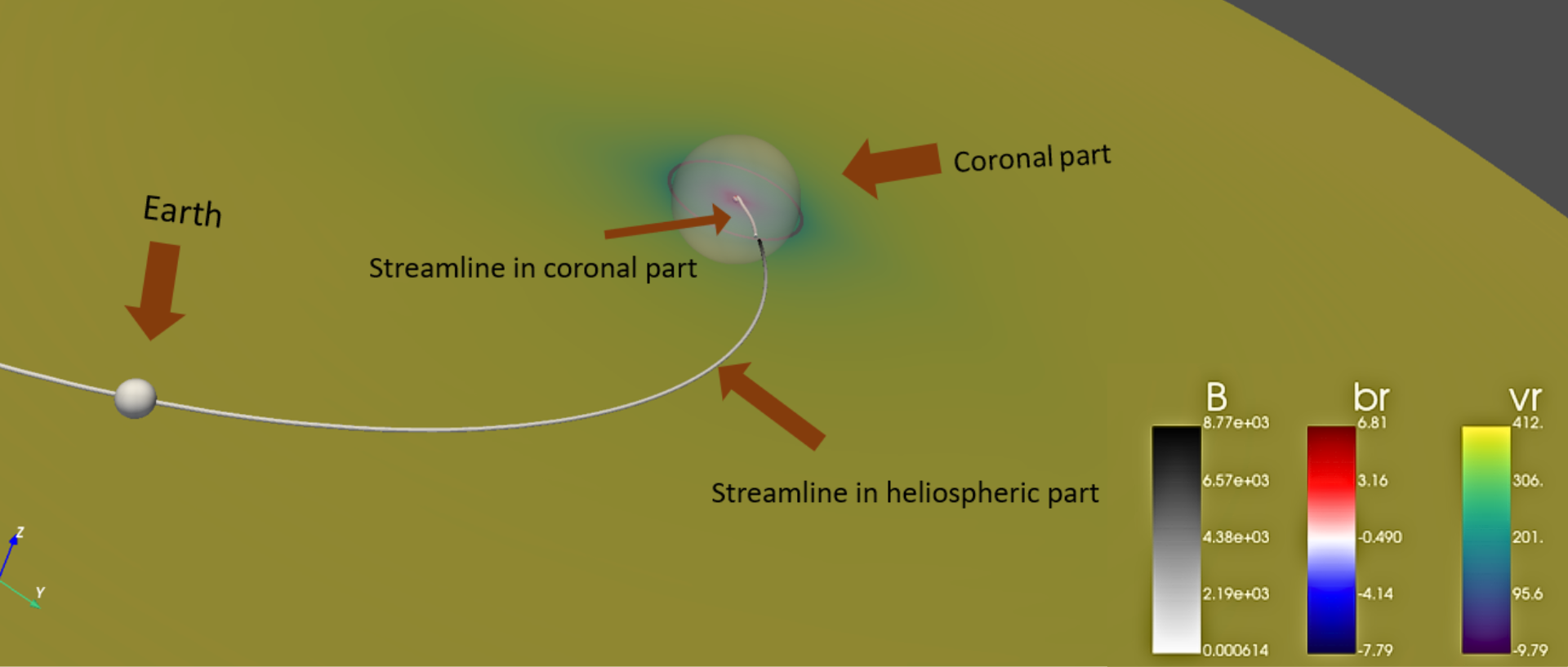}
    \caption{Example of magnetic connectivity computed between the Earth and the Sun using COCONUT and EUHFORIA models. The Earth is depicted as a gray sphere to show the initial seed of the magnetic field line. Then, we have a first field line computed in EUHFORIA in the heliospheric part (shown as a white flux tube). The interface between the coronal and heliospheric parts (COCONUT and EUHFORIA) is shown via a transparent sphere at 0.1 AU. Finally, inside the coronal part, we show the second magnetic field line computed using COCONUT results (white flux tube inside the transparent sphere).}
    \label{fig:field line1}
\end{figure*}

We can start tracing the magnetic field lines once the two models are in the same reference frame. We use the PyVista Python package \citep{Sullivan2019}, which reads in the COCONUT and EUHFORIA 3D datasets and traces streamlines using a Runge-Kutta 2 integrator. The streamlines of a magnetic vector field are what we call magnetic field lines. We start from the position of the Earth, which is always located in the heliospheric part, so the data from EUHFORIA is used first. From the magnetic field computed by EUHFORIA, we trace the field lines down to $\sim$0.1~AU. The endpoint of the field line in the heliospheric part for EUHFORIA gives the starting seed for the field line in the coronal part for COCONUT. We finally trace the second magnetic field line from this interface point until the inner boundary of COCONUT, which corresponds to the inner corona close to the solar surface. This second streamline's endpoint is then Earth's connectivity point on the solar surface. 
The resulting magnetic connectivity can be seen in Figure~\ref{fig:field line1}. The Earth is depicted as a gray sphere to show the initial seed of the magnetic field line. Then, we have a first field line computed in EUHFORIA in the heliospheric part (shown as a white flux tube). The interface between the coronal and heliospheric parts (COCONUT and EUHFORIA) is shown via a transparent sphere at 0.1~AU. Finally, inside the coronal part, we show the second magnetic field line computed using COCONUT results (white flux tube inside the transparent sphere). 
To sum up, for each date, we can provide the magnetic connectivity point at the surface of the Sun that is magnetically connected to the Earth. 
This procedure is used here only for the connectivity between the Earth and the Sun, but the same principle could be applied to other satellites. 

A complementary approach is to use the velocity streamlines as another indicator of connectivity in addition to the magnetic field lines: because of the frozen-in theorem, magnetic and velocity streamlines should coincide most of the time. However, we cannot be certain that this is always the case: in the case of magnetic reconnection in particular, magnetic field lines may become discontinuous, but not velocity streamlines, hence simplifying the complexity of certain events. The use of velocity streamlines will be examined more thoroughly in a future study.

Because we work with realistic field lines, they are not always continuous. This can be caused by numerical reconnection in the current sheet due to the resolution or by shocks and discontinuities that form naturally in the numerical domain. We have developed specific procedures to exclude these exceptions and only compute the field lines that extend between the two models. This means that we exclude field lines in EUHFORIA that do not reach the interface at $\sim$0.1~AU, and we also exclude field lines in COCONUT that do not reach the inner boundary condition at one solar radius. 

We perform an event based study to test the accuracy of our connectivity estimates. The events and how they were chosen can be seen in Section~\ref{sec:Caseselection}. To assess the robustness of our results, we have computed uncertainties associated with each event. First, we have taken into account a temporal uncertainty: this comes from the fact that the solar wind takes on average four days to travel from the Sun to the Earth, but this is a statistical approximation obtained for High-Speed Streams (HSSs) produced by Coronal Holes (CHs) \citep{Vrsnak2007}. In practice, the solar wind speed can vary along the trajectory and thus impact the propagation time \citep{koukras2022}. To take this into account, when we are trying to derive the magnetic connectivity at an event seen at Earth for a specific date, we average the connectivity points we find over the past few days before the event to get an idea of the temporal evolution of the magnetic connectivity over the propagation phase of the solar wind. However, we specify that this temporal variation is made just by rotation of the simulation, so by assuming that the magnetic field configuration did not vary over the studied period (which is not always the case at maximum of activity). Second, we also take into account spatial uncertainty: as explained before, the magnetic field lines can be discontinuous, or they can also be very divergent locally, which makes it risky to rely on a single field line for the magnetic connectivity estimate. This spatial uncertainty also includes the possible deviation of a few degrees induced by the neglect of the effect of the rotation. To go beyond, we thus compute several field lines for the Earth connectivity, each starting from a seed point separated from Earth by an angle between 5 and 10 degrees in latitude and/or longitude. Ultimately, we use 9 points around the Earth's position and thus compute between 1 and 9 field lines, depending on their continuity. 
Thus, we can ultimately get several dozens of points for the magnetic connectivity estimate. We take the mean value of the points' positions and compute the standard deviation to provide a magnetic connectivity zone instead of a magnetic connectivity point. We have implemented a rule that if the difference in latitude exceeds 20 degrees, the zone is split into two different sub-zones. 

\subsection{Alternative way of computing connectivity for validation}
\label{sec:Alternative ways}
To validate our magnetic connectivity estimation, we will compare it with other studies such as \cite{Reiss2021} and \cite{koukras2022}. The first study is purely observational, focusing on an HSS case where the connected coronal hole could be easily identified (as described in more details in the next section). The second study uses a modified back-mapping method, replacing the ballistic part with a two-part Parker solution approximation. In contrast, the PFSS part is based on GONG magnetograms.
We will also compare our MHD connectivity estimate with a more traditional one based on the usual back-mapping method. We will describe this method as implemented here, using the two steps of ballistic mapping and magnetic mapping.

Ballistic mapping is where the solar wind is traced from a point in the inner heliosphere to the source surface using the Parker spiral corresponding to the in situ measured solar wind speed $v_r$ and assuming it is constant with distance \citep{Krieger1973, Peleikis2017}. For a spiral of heliographic colatitude $\theta$, the heliographic longitude $\varphi(r)$ of the spiral as a function of heliocentric distance ($r$) relative to its starting longitude is given by:
\begin{equation}
    \varphi(r) = -\frac{\Omega_\odot r \sin(\theta)}{v_r}
\end{equation}
where $\Omega_\odot$ is the angular velocity of the Sun. Strictly speaking, it should  be $(r-r_{ss})$ where $r_{ss}$ is the source surface height to be self-consistent with the PFSS condition that field lines are radial at the outer boundary, but we assume this correction to be small enough so that we can neglect it.

The second concept used together with the Parker spiral for validation is the Potential Field Source Surface (PFSS) extrapolation \citep{Altschuler1969, Schatten1969, schrijver2003}. This model focuses on the structure of the coronal magnetic field in its minimum energy state.
This simplistic model assumes that the coronal magnetic field is potential, i.e., the rotation of $\mathbf{B}$ is zero, so it cannot capture twisted structures. The PFSS extrapolation uses a photospheric magnetogram as the inner boundary condition for the magnetic field potential, which satisfies the Laplace equation. The magnetic field is assumed to be purely radial starting from the source surface (usually around 2.5 solar radii, see \cite{Hoeksema1983}).
This method is very efficient in estimating the magnetic field geometry in the solar atmosphere and is therefore commonly used in space weather forecasting \citep{Lee2011, Pomoell2018}.
To implement this technique, the Python package \textit{pfsspy} was used, which is an open source code fully integrated into the\textit{sunkit-magex} package\footnote{ More information can be found here: \url{https://github.com/sunpy/sunkit-magex}.} within \textit{sunpy}. \citep{Stansby2020}. For the latter calculations, a 3D grid, with grid points (720,360,50) is used. For each case, a source surface radius of 2 was chosen by comparing the HCS from PFSS with the one simulated with the MHD model, and then by selecting the best match.

\begin{figure}[!h]
    \centering
    \includegraphics[width=\hsize]{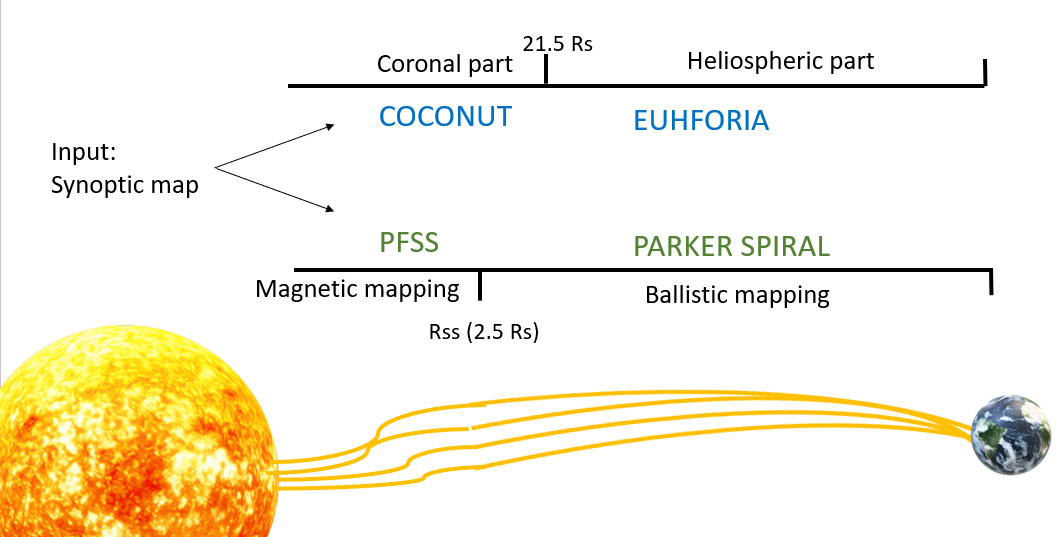}
    \caption{Illustration of the two different chains of models used to compute the magnetic connectivity in this paper. The first chain is our combination of MHD models, COCONUT and EUHFORIA. The second one is the traditional back-mapping chain with the PFSS extrapolation coupled with the ballistic mapping.}
    \label{fig:models}
\end{figure}

A summary of the two different methods used for the magnetic connectivity estimations of this paper can be found in Figure~\ref{fig:models}. 

\section{Description of the validation cases}
\label{sec:Caseselection}
To quantify the quality of our magnetic connectivity estimation, we need reference cases to validate our results against observations. However, magnetic connectivity is challenging due to the lack of direct measurements. The closest we have to direct observations is the ion charge and elemental measurements \citep{Landi2012}. Indeed, composition tends to be frozen in the solar wind, hence we can, in theory, connect spectroscopic and in situ measurements with the same ion charge states (oxygen or carbon ratio). \citet{Baker2023} and \citet{Yardley2024} are two examples where remote sensing and in situ data were combined to obtain the magnetic connections. However, this technique has not yet been applied systematically, which means that very few validation cases have been identified unambiguously. Therefore, the second-closest method we have to determine the magnetic connectivity between the Sun and the Earth is to use High-Speed Stream (HSS) events.  

HSSs are fast solar wind streams detected at Earth \citep{Snyder1963, Grandin2019}. They are very convenient to study because their identification can often be unambiguous (shock and jump from 400 km/s to 700 km/s) and they can easily be connected to the solar surface. Indeed, the sources of the fast solar wind are currently better understood than the ones from the slow solar wind \citep{Cranmer2017} and have been determined to be the central region of coronal holes \citep{Krieger1973, McComas2000, Kohl2006}. Coronal holes (CHs) are low-density patches in the solar corona due to open magnetic field lines, resulting in a darker region in Extreme Ultra-Violet (EUV), which can also be easily identified \citep{Zirker1977}. By assuming that the magnetic field lines are frozen in the solar wind, we can use these events to validate the magnetic connectivity.

\begin{figure*}[!h]
    \centering
    \includegraphics[width=\hsize]{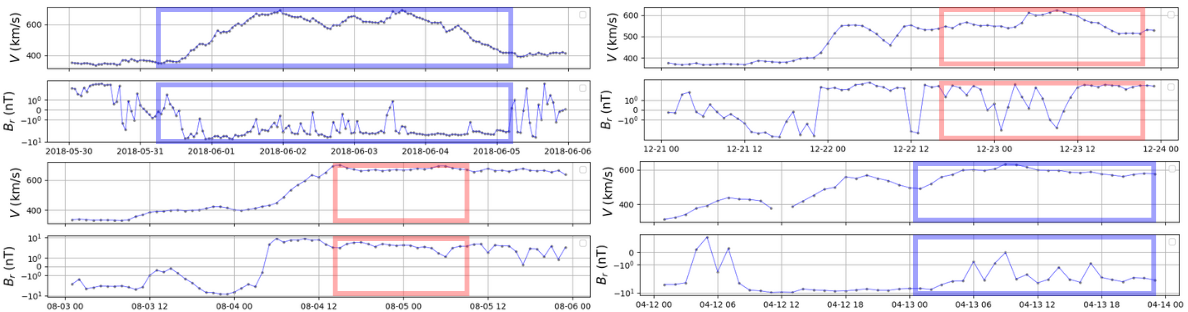}
    \caption{In situ data for events 1 to 4 (left to right, then top to bottom). For each event, we show the mean velocity of the solar wind in km/s (top subpanel) and the radial magnetic field in nanoTesla (bottom subpanel). The ACE data, measured in situ, helps identify the high-speed stream corresponding to the coronal hole event and indicates the polarity at Earth. The red and blue squares indicate the specific period of the event where a positive and negative polarity is expected, respectfully.}
    \label{fig:in-situ}
\end{figure*}

\begin{figure}[!h]
    \centering
    \includegraphics[width=\hsize]{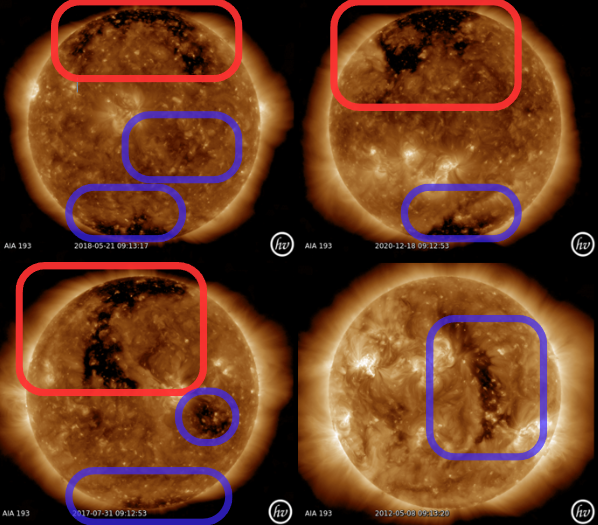}
    \caption{ AIA (193\text{\AA}) images from SDO for events 1 to 4 (left to right, then top to bottom). The coronal holes are the darker regions visible and marked in the EUV wavelengths. The red and blue squares indicate the specific coronal hole where a positive and negative polarity is expected, respectfully. The figures are made with Helioviewer \citep{helioviewer}. More information about the events can be found in Table~\ref{tab:Dates}.}
    \label{fig:AIA}
\end{figure}

In particular, we use four cases previously identified and studied in the literature (see Figure~\ref{fig:in-situ}). The coronal holes that can be a possible source can be found in Figure~\ref{fig:AIA}. The first one is from \citet{Reiss2021} and corresponds to an HSS reaching Earth on the 1st of June 2018 (event 1, top left panel). The other three cases are from \citet{koukras2022} and correspond to various HSSs on the 22nd of December 2020 (event 2, top right panel), the 4th of August 2018 (event 3, bottom left panel) and the 13th of April 2012 (event 4, bottom right panel). 
In the selected cases, only pure fast solar wind events were chosen to avoid the perturbations induced by compression regions. 
We display the in situ measurements of the mean solar wind velocity in km/s (top subpanel) and the radial magnetic field in nanoTesla (bottom subpanel) for all events. These data come from the satellite ACE (Advanced Composition Explorer) at Lagrange point L1. Three cases are during a solar minimum (events 1, 2 and 3) and one during a solar maximum (event 4). The solar maximum case is expected to be more challenging because solar activity can affect the accuracy of the magnetic connectivity. 

The radial magnetic field component indicates the polarity of the interplanetary magnetic field. A positive value corresponds to a positive polarity (field line going away from the Sun), while a negative value indicates a negative polarity (field line going towards the Sun). Please note that ACE does not provide directly $B_r$, but instead, the GSE frame coordinates\footnote{More information about the various ACE coordinate systems can be found here: \url{https://izw1.caltech.edu/ACE/ASC/coordinate_systems.html}.}, so we assumed that $B_r \sim -B_{GSE,x}$. The implications of the polarity will be discussed in more detail in Section \ref{sec:Polarity}. 

The in situ data for our events can be seen in Figure~\ref{fig:in-situ}. 
For the first event in 2018 (top left panel), we show the data between 31/05/2018 and 05/06/2018. The polarity for that event is clearly negative (marked by a blue rectangle).
For the second event in 2020 (top right panel), we see that the polarity changes a lot between 22-23/12/2020, but is primarily positive (marked by a red rectangle).
For the third event in 2017 (bottom left panel), we can see that the polarity goes from negative to positive and stays positive during the HSS.
For the last event in 2012 during a solar maximum (bottom right panel), the overall magnetic polarity is negative.

Since we also have access to the CHs polarity at the surface of the Sun, this gives us a tool to pinpoint more precisely which CH is responsible for the observed HSS of the same polarity. That way, we can complete the in situ data from Figure~\ref{fig:in-situ} with remote-sensing data in Figure~\ref{fig:AIA}. This figure displays the Sun in EUV (193$\AA$ from SDO/AIA), which allows us to see the coronal holes responsible for the fast solar wind as darker regions. We have marked the largest CHs with rectangles whose color is associated with their polarity (red for positive, blue for negative). Going back to what we described earlier for the in situ data, we can then assume that for event 1 (top left panel), the HSS observed at Earth originates from the central CH (which is what was found in \cite{Reiss2021}) or the southern CH. Similarly, event 2 originates from the northern CH, event 3 also originates from the northern CH, and finally event 4 originates from the equatorial CH. 

With the HSS events well identified at Earth, we perform the corresponding coronal and heliospheric numerical simulations to reproduce them. To do so, we use a magnetic map as input that covers the period before the event. Indeed, since solar wind can take 2 to 5 days on average to reach the Earth, we cannot select a single magnetic map at a specific date to reproduce the exact solar configuration that triggered the HSS. Instead, we use HMI Carrington maps that accumulate observations over the full Carrington rotation for both the MHD model and the back-mapping method. These maps have also been shown to give the best results for the COCONUT model \citep{Perri2023}. The most realistic way to model the magnetic connectivity would be, of course, to have a time-dependent model \citep{Lionello2023}. Still, this feature is not available yet for the COCONUT and EUHFORIA models and is thus outside the scope of this paper. 

\begin{table*}[!h]
    \centering
    \begin{tabular}{p{0.75cm} |  p{2.5cm}  p{2.5cm} | p{2.5cm} p{2.5cm} | p{2.25cm} | p{3cm}}
Event    & \multicolumn{2}{c|}{Observed at the Earth} & \multicolumn{2}{c|}{Period of magnetogram} &  Solar minimum / maximum & Corresponding study \\ 
 & Start Date & End Date & Start Date & End Date &  &  \\
 \hline
    1 &  31/05/2018 10:08 & 05/06/2018 10:21 &16/05/2018 10:57 & 12/06/2018 05:55 & Minimum & \citet{Reiss2021}  \\ 
   2  &     22/12/2020 17:59 & 23/12/2020 15:10&  28/11/2020 14:46 & 25/12/2020 22:31  & Minimum &  \citet{koukras2022}\\
   3  &     04/08/2017 13:19 & 05/08/2017 10:11 & 20/07/2017 06:08 & 16/08/2017 11:27   & Minimum & \citet{koukras2022}\\
   4 &     13/04/2012 04:07 &  13/04/2012 23:13& 31/03/2012 21:25 & 28/04/2012 03:48 & Maximum & \citet{koukras2022}\\
    \end{tabular}
    \caption{Arrival times of solar wind on Earth and period covered by magnetogram at the Sun. These events are used to calculate the magnetic connectivity between the Earth and the surface of the Sun for a given period. The telescope used for all the events is SDO/HMI. The events are taken from two different studies, from \citet{koukras2022} and \citet{Reiss2021}.}
    \label{tab:Dates}
\end{table*}

The relevant information about the validation events, including the corresponding arrival of the solar wind at Earth and the period covered by the magnetic map of the Sun for the four cases studied, can be found in Table~\ref{tab:Dates}.

\section{Validation of the magnetic connectivity}
\label{sec:Connectivity}

\begin{figure*}[!h]
    \centering
    \includegraphics[width=\hsize]{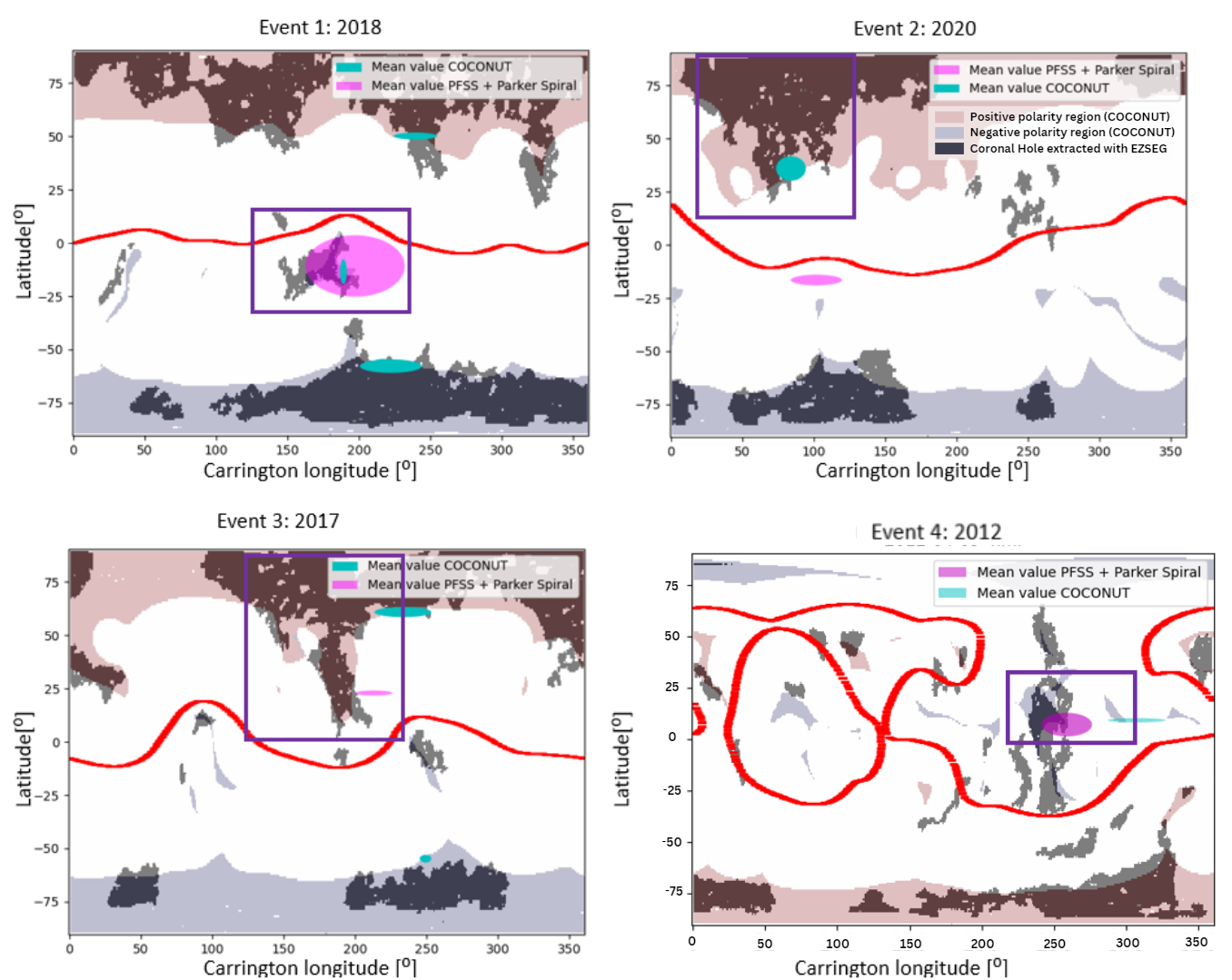}
    \caption{Validation of the MHD connectivity estimation for the four cases selected. We show the location at the surface of the Sun, which is most likely to be connected to Earth during the corresponding HSS event. The MHD connectivity estimate is shown in bright blue, the two-step ballistic estimate is shown in pink, and the magnetic connectivity estimate from other studies (\cite{Reiss2021} for event 1 and \cite{koukras2022} for other events) is shown with a purple rectangle. The area covered by the magnetic connectivity estimate represents the spatial and temporal uncertainties. We also show the HCS (red line) and coronal holes to explain these results better. The red and dark blue patches indicate respectively the positive and negative polarity regions of open magnetic field lines in the COCONUT simulation, while the grey areas indicate the CHs extracted from the SDO Carrington EUV maps using the EZSEG algorithm for Events 1-3 \citep{Caplan2016}. 
    }
    \label{fig:connection_samen}
\end{figure*}

We now compare our magnetic connectivity MHD estimations with results from previous studies and the traditional two-step ballistic mapping. The results for all four events are plotted together in Figure~\ref{fig:connection_samen}: event 1 (2018) is on the top left, event 2 (2020) is on the top right, event 3 (2017) is on the bottom left and event 4 (2012) on the bottom right. We show the location at the surface of the Sun which is the most likely to be connected to Earth during the corresponding HSS event. The MHD connectivity estimate is shown in bright blue, the two-step ballistic estimate is shown in pink, and the magnetic connectivity estimate from other studies (\cite{Reiss2021} for event 1 and \cite{koukras2022} for other events) is shown with a purple rectangle. The area covered by the connectivity estimate represents the spatial and temporal uncertainties. We also show the HCS (red line) and coronal holes to explain these results better. The red and dark blue patches indicate, respectively, the positive and negative polarity regions of open magnetic field lines in the COCONUT simulation, while the grey areas indicate the CHs extracted from the SDO Carrington EUV maps using the EZSEG algorithm \citep{Caplan2016}. More details about this method can be found in \cite{Perri2023}.

For the first event (top left panel), the simulation shows three potential MHD connectivity points. One magnetic connection is in the northern coronal hole, with a positive polarity; another is connected to a negative polarity southern coronal hole, and the last is in an equatorial coronal hole, which also has a negative polarity.
The two-step ballistic mapping method yields a mean magnetic connectivity point located around the equatorial coronal hole, which has a negative polarity. 

In the simulation, the Earth is mainly located beneath the HCS, so it is more likely that the magnetic connection is with the equatorial or the southern CH, which have the same polarity as the in-situ data and the two-step ballistic mapping. For this event, it was previously determined that the correct magnetic connection is with the equatorial coronal hole (see \cite{Reiss2021} for more details). The results of EUHFORIA and COCONUT also support this outcome but with more uncertainty than the back-mapping method. This could be attributed to the simulation parameters (as the equatorial CH is not that big) or the complexity of the field line configurations (which could explain the incursion of a positive polarity patch in this region). A summary of the polarity for all cases can be found in Table~\ref{tab:results_COCONUT_magneticpol}. 

If we then look at event 2 (top right panel), we see that the MHD simulation gives a result around the northern coronal hole with a positive polarity, similar to the result obtained in the study of \citet{koukras2022}. However, the two-step ballistic method gives a mean value around the equator with a negative polarity. In this case, the result obtained with COCONUT can be assumed to be correct because it has the same magnetic polarity as the in-situ data. This shows that MHD simulations can give an accurate estimate of the magnetic connectivity even in complex cases, as here, the event corresponds to a very patchy coronal hole located close to active regions (i.e., closed field lines very close to open field lines). 

\begin{table*}[!h]
    \centering
    \begin{tabular}{c||c||c||c}
    & \multicolumn{1}{c||}{COCONUT + EUHFORIA} & \multicolumn{1}{c||}{Back mapping method}  & \multicolumn{1}{c||}{Ground Source }\\ 
    \hline 
    \hline 
      Event  & Magnetic Polarity & Magnetic Polarity &  Magnetic Polarity  \\
      \hline 
      1 &  Positive  &  Negative &   Negative  \\
        & Negative  & & \\
        &  Negative  & &   \\
        \hdashline
      2 &  Positive&  Negative &  Positive \\
      \hdashline
      3 &  Positive  &  Positive &  Positive\\
        &  Negative & &  \\ 
        &  Negative & & \\
     \hdashline
      4 &  Negative & Negative &  Negative\\
    \end{tabular}
    \caption{ A summary of the magnetic polarity of the found results from COCONUT and EUHFORIA and the two-step ballistic mapping method. These results in numerical form can be found in Table~\ref{tab:results_COCONUT}}. The right column gives the  average magnetic connectivity of the in-situ data 
    \label{tab:results_COCONUT_magneticpol}
\end{table*}


For event 3 (bottom left panel), the MHD simulations produced three different results: a magnetic connection point at a southern CH with a negative polarity, at a northern CH, with a positive polarity, and at an equatorial CH, also with a negative polarity. This last one is difficult to see because of the minor uncertainties. The back-mapping method also gave a result around the CH in the north, with a positive polarity, consistent with the result obtained by \cite{koukras2022}. Therefore, it is more likely that the actual magnetic connection is at the northern coronal hole, where both the simulation and the in-situ data have a positive polarity. 


Finally, for event 4 (bottom right panel), we can see that all the results of the different approaches are located around the equator, with both a negative polarity. The HCS from the simulation (red line) is way more complex and less flat than in the previous cases, which seems to provide more precise results. The results obtained with COCONUT and EUHFORIA are located in a CH close to the one from \cite{koukras2022}. We do not have the observations needed to confirm or refute which one is the correct CH (\cite{koukras2022} assumed it was the largest one).  


The coordinates of the connectivity estimates derived using the MHD models and the two-step ballistic mapping are summarised in Table~\ref{tab:results_COCONUT}. Latitudes are in degrees, longitudes in degrees in the Carrington frame. The table's last column indicates the overlap between the results obtained from COCONUT and EUHFORIA and those obtained by \citet{Reiss2021} and \citet{koukras2022}. The percentage of overlap was calculated by dividing the total number of magnetic connections to the right CH throughout the entire period by the total number of connections obtained during that period.

\begin{table*}[!h]
    \centering
    \begin{tabular}{c||c|c||c|c||c}
    & \multicolumn{2}{c||}{COCONUT + EUHFORIA} & \multicolumn{2}{c||}{Back mapping method} & Overlap [$\%$] \\ 
    \cline{2-6}
    \hline 
    \hline 
      Event  &Mean longitude [$^{\circ}$] & Mean Latitude [$^{\circ}$] & Mean longitude [$^{\circ}$] & Mean Latitude  [$^{\circ}$] \\
      1 & 239.36 $\pm$ 15.25 & 50.18 $ \pm$ 1.65 & 197.21 $\pm$ 34.74  & -10.80 $\pm$ 14.42 & 19.45 \\
        & 189.02 $\pm$ 2.28 & -13.45 $\pm$ 5.88 & & \\
        & 222.20 $\pm$ 21.66 & -57.79 $\pm$ 3.18 & &   \\
        \hdashline
      2 &83.98 $\pm$ 10.4& 36.01 $\pm$5.79 & 115.40 $\pm$ 14.71 & -17.33 $\pm$1.42 & 100 \\
      \hdashline
      3 & 233.38 $\pm$ 19.89  & 60.90 $\pm$ 2.37& 213.01 $\pm$ 13.03 & 22.93 $\pm$	1.15 & 45.03 \\
        & 252.65 $\pm$ 0.21 & -19.45 $\pm$	0.20 & &  \\ 
        & 249.21 $\pm$ 4.09 & -54.80 $\pm$ 1.81 & & \\
     \hdashline
      4 & 307.1 $\pm$ 20.12	& 9.26 $\pm$ 1.05& 258.87 $\pm$ 17.40 & 7.12 $\pm$	5.73 & 100\\
    \end{tabular}
    \caption{Mean values of the magnetic connectivity locations for the four events studied (latitudes in degrees, longitudes in degrees in the Carrington frame). The first column shows the estimates obtained from COCONUT and EUHFORIA, while the second column shows the magnetic connectivity obtained from the two-step ballistic mapping method. The final column indicates the percentage of agreement between data and simulations.}
    \label{tab:results_COCONUT}
\end{table*}

What is very surprising here is that the most difficult cases (event 2 because of patchy CH and event 4 because of maximum activity) are the ones where the connectivity is better estimated. Events 1 and 3 yield some overlap (respectively 19 and 45\%), so the MHD connectivity is only partly correct. This can be explained for event 1 by the small opening of the equatorial CH, but is more difficult to explain for event 3. To understand these results further, we will look more precisely at the polarity variations in the next section. 

\section{Discussion about polarity estimation}
\label{sec:Polarity}

To understand why certain events provide good magnetic connectivity while others do not, we can explore the magnetic configuration on a larger scale and compute the polarity seen at Earth over the full Carrington rotation in our simulations. Indeed, if the in situ polarity is not correct at the date of the event, we cannot recover the proper CH connectivity. It then tells us that the cause of the error is more complex and linked to a mistake in the HCS position. The work of \cite{Badman2022} inspired this analysis, where a similar study was applied to solar wind simulations for comparison with PSP data.

\begin{figure*}[h]
    \centering
    \includegraphics[width=0.3\hsize]{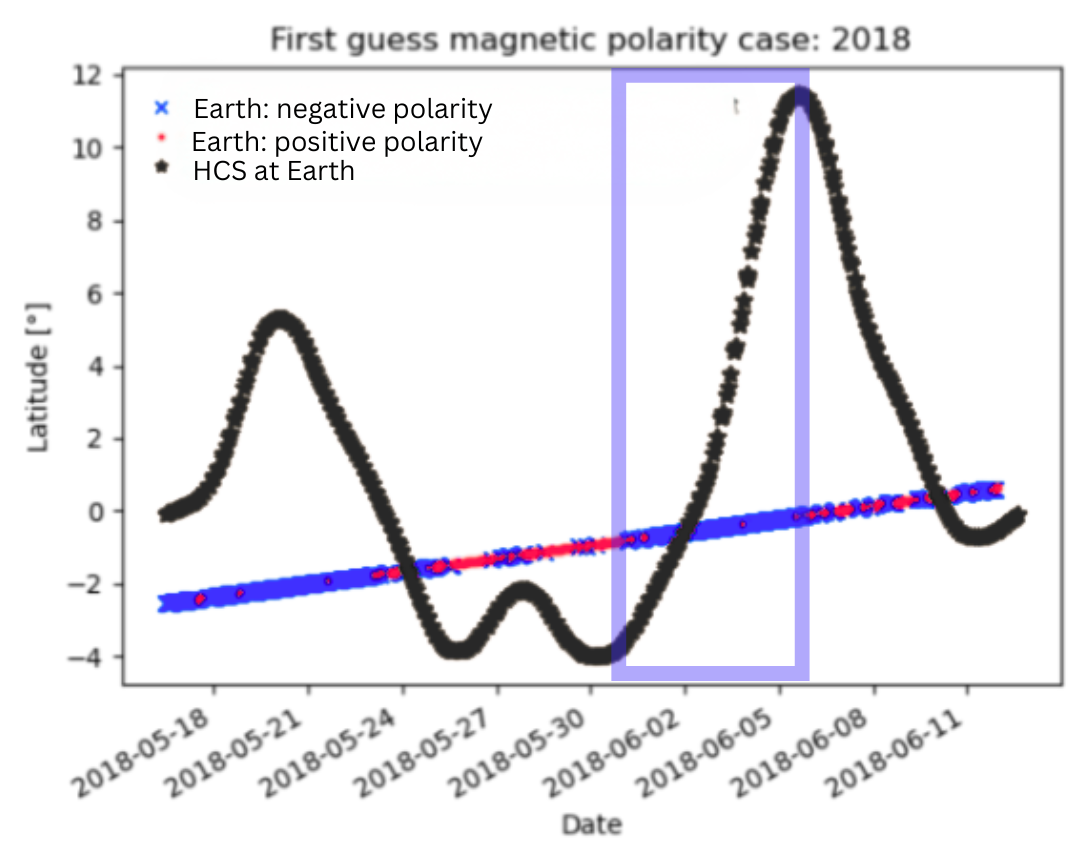}
    \includegraphics[width=0.3\hsize]{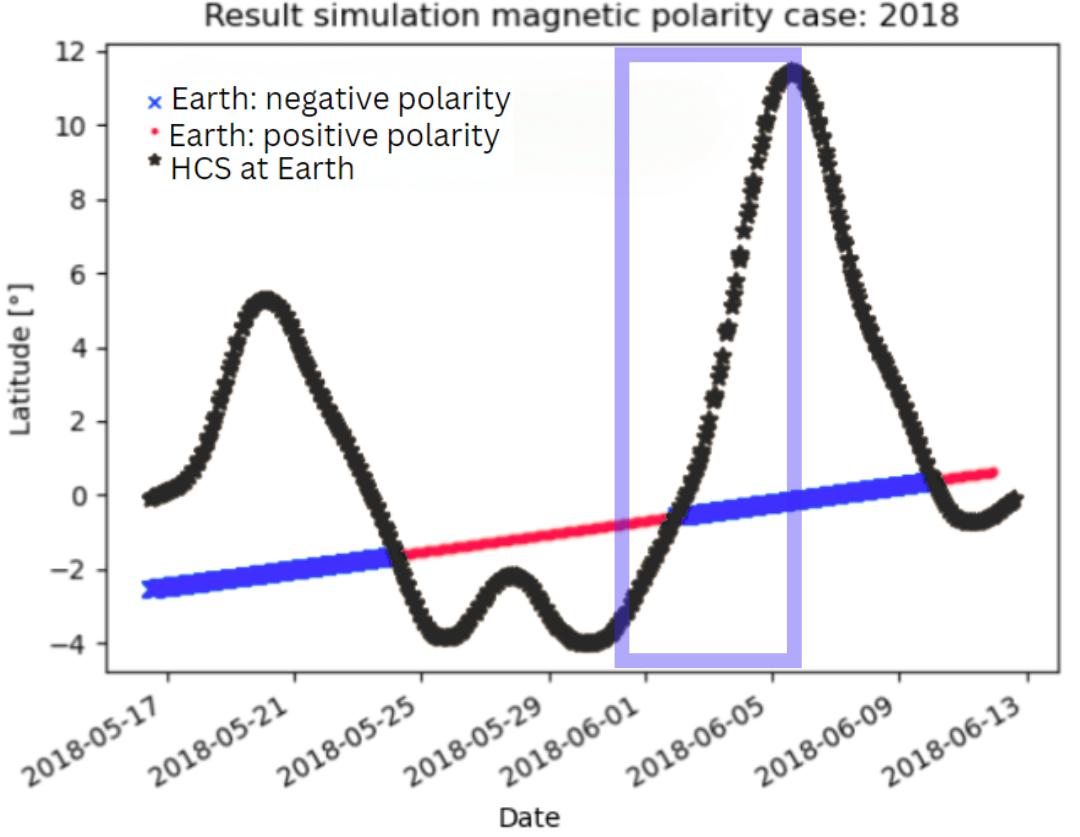}
    \includegraphics[width=0.3\hsize]{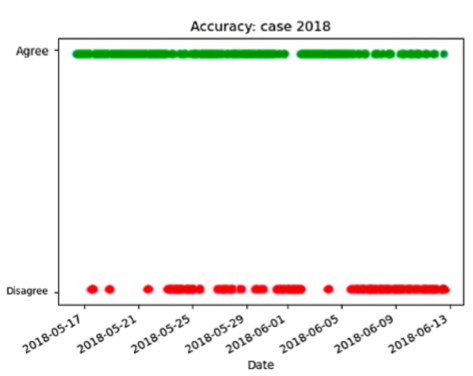}

    \includegraphics[width=0.3\hsize]{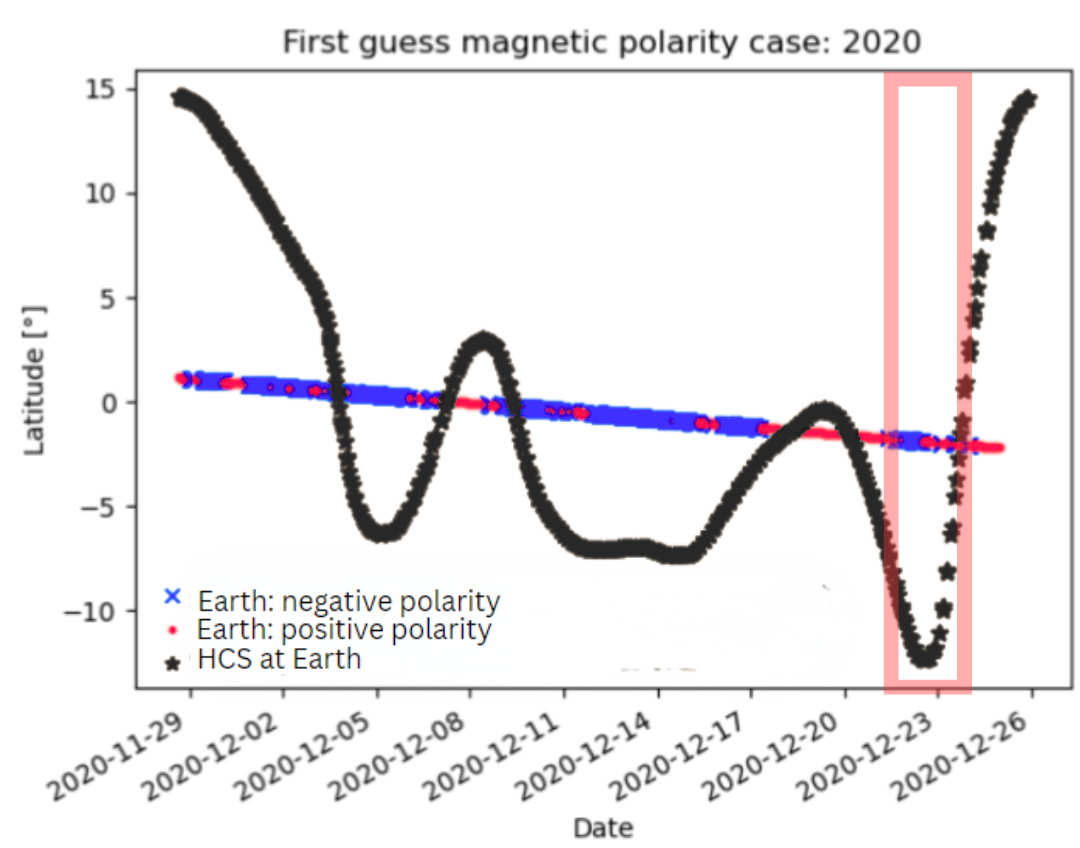}
    \includegraphics[width=0.3\hsize]{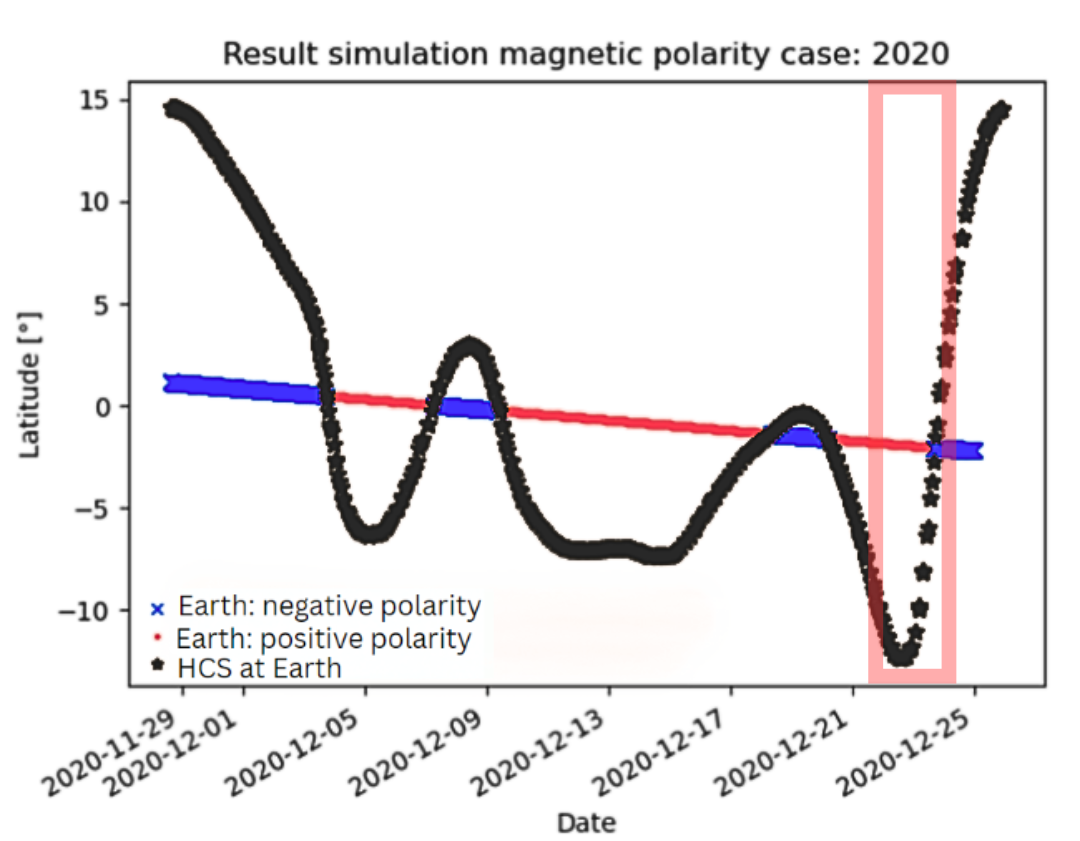}
    \includegraphics[width=0.3\hsize]{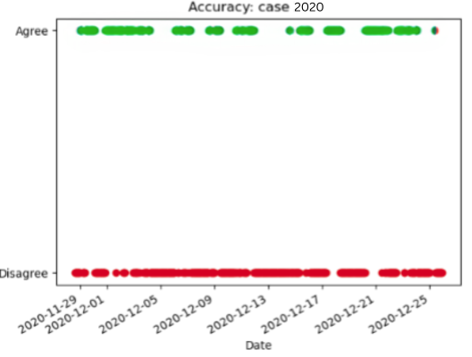}

    \includegraphics[width=0.3\hsize]{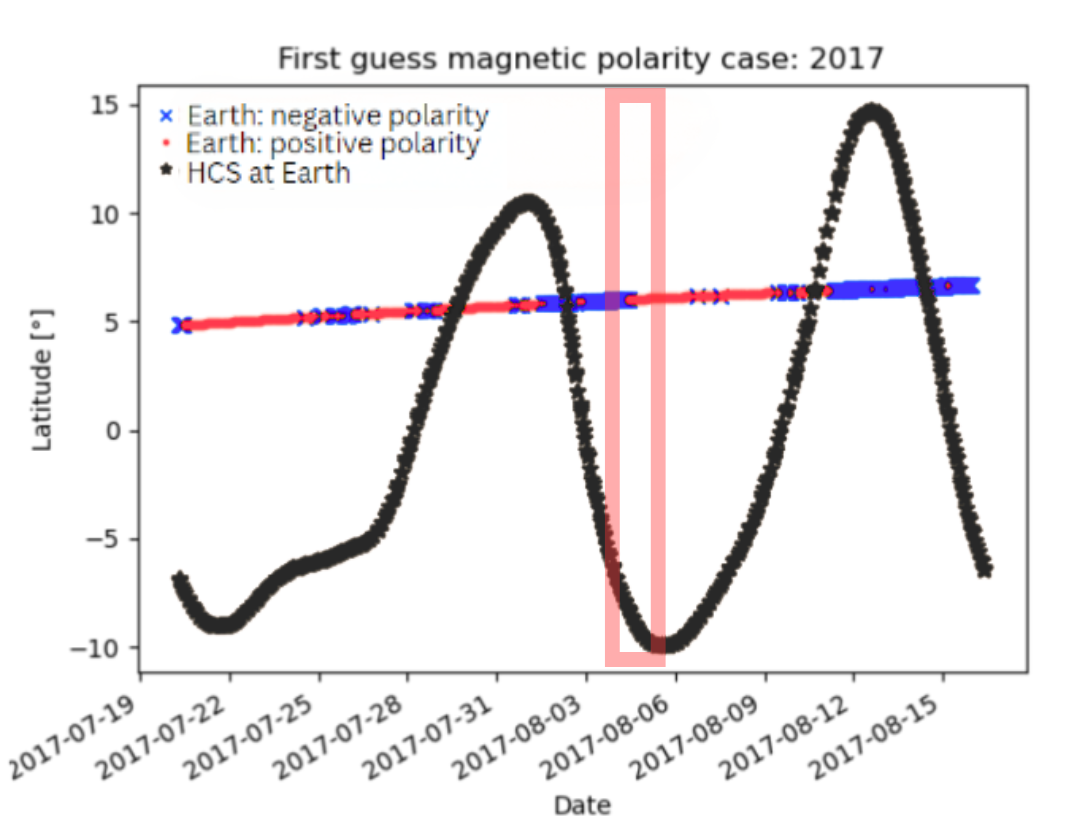}
    \includegraphics[width=0.3\hsize]{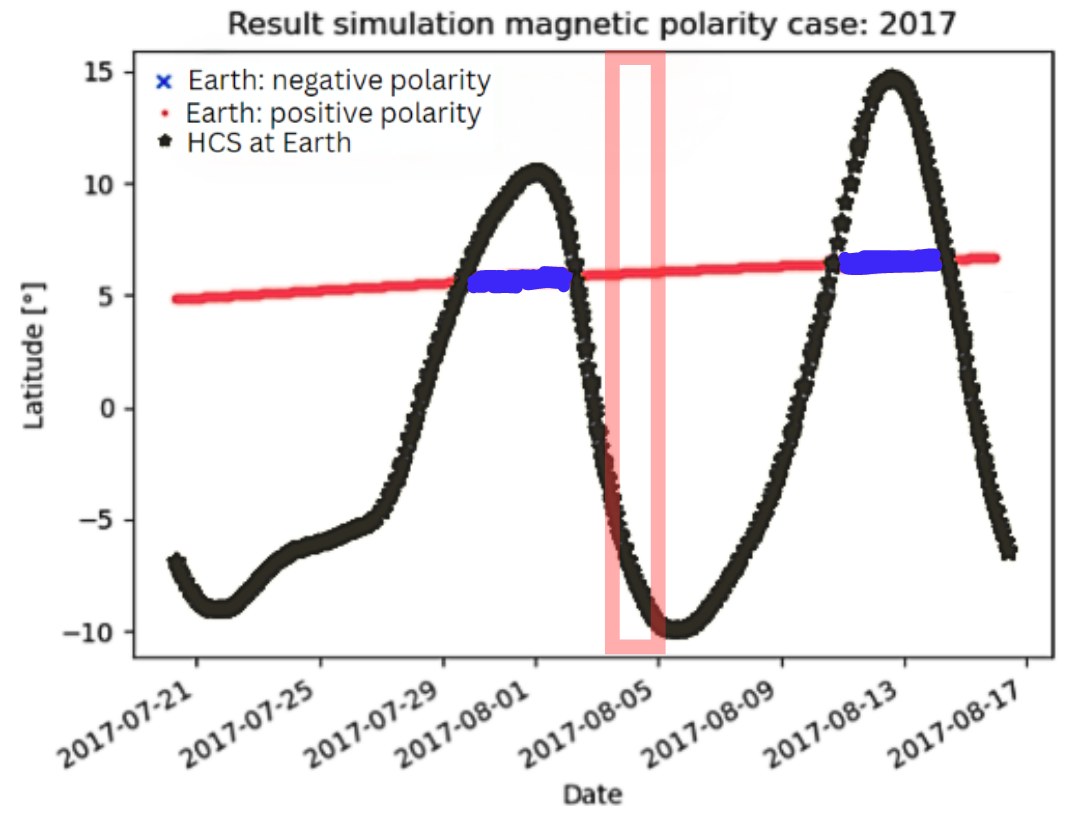}
    \includegraphics[width=0.3\hsize]{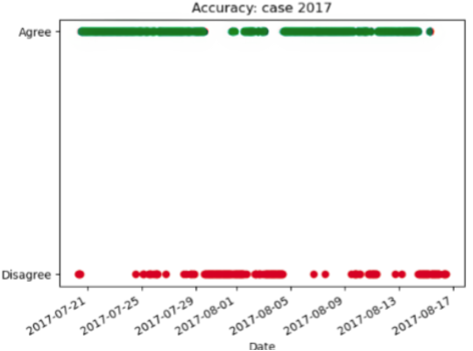}

      \includegraphics[width=0.3\hsize]{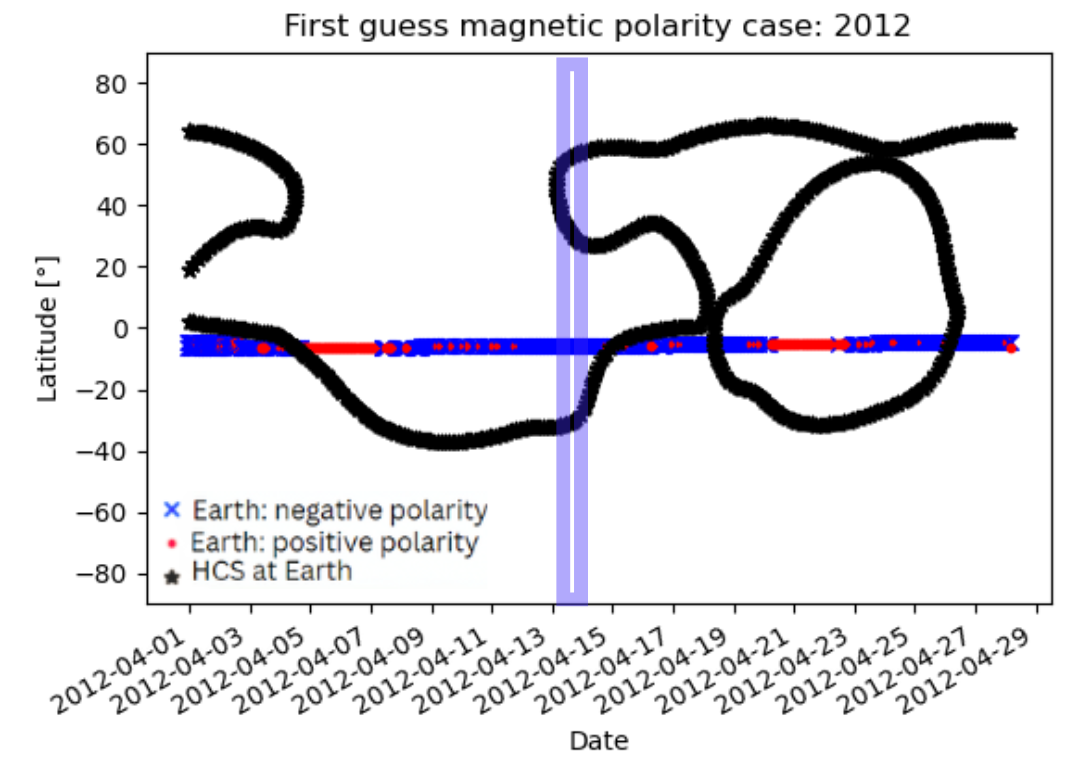}
    \includegraphics[width=0.3\hsize]{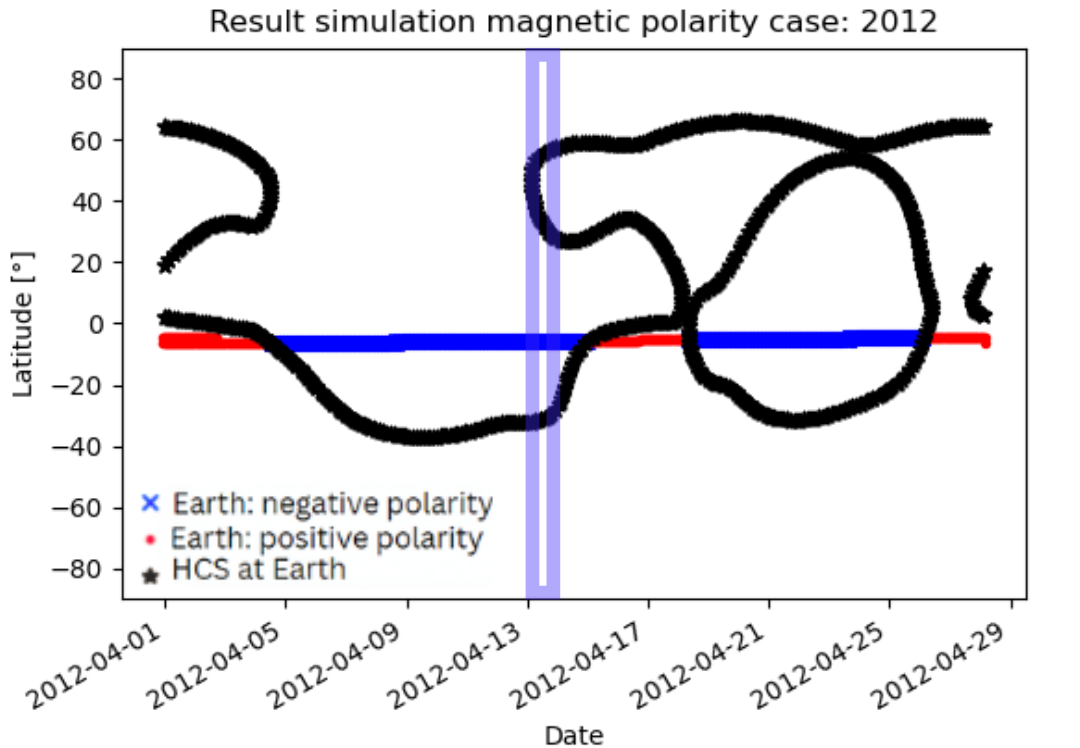}
    \includegraphics[width=0.3\hsize]{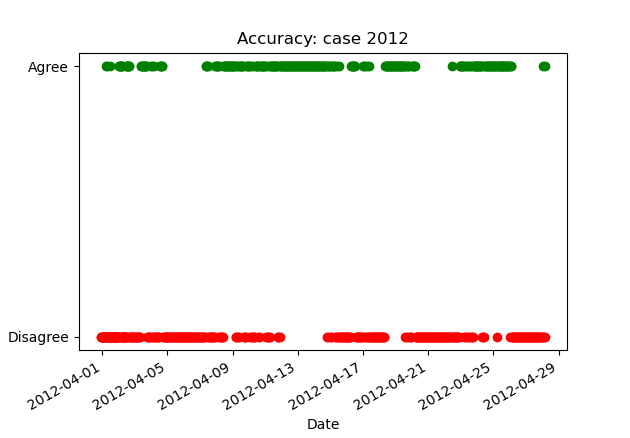}
    \caption{Validation of the global Earth magnetic connectivity using in situ polarity. Each line corresponds to one of the studied events (event 1 for line 1, and so forth). For the left column, we plot the Earth's latitude for each date over the simulated period and indicate the corresponding in situ polarity measured by ACE (blue for negative, red for positive). We put this in perspective with the simulated HCS at Earth's orbit (black line). The middle column is essentially the same plot, except that this time, the polarity at Earth positions comes from the MHD simulations. Finally, the right column quantifies the agreement between in situ and simulated polarity for each date (in green when there is agreement, in red otherwise). These plots were inspired by the work of \cite{Badman2022}. 
    }
    \label{fig:exp_samen}
\end{figure*}

The corresponding analysis can be found in Figure \ref{fig:exp_samen}. 
Each line corresponds to one of the studied events (event 1 for line 1, and so forth). For the left column, we plot the Earth's latitude for each date over the simulated time and indicate the corresponding in situ polarity measured by ACE (blue for negative, red for positive).
The latitude of the Earth is determined using the SpiceyPy Python package \citep{andrew2020}. Each date is obtained by rotating the steady-state simulation to the corresponding Carrington longitude. 
The in situ data are the same as shown in Figure~\ref{fig:in-situ}. 
We put this in perspective with the simulated HCS at Earth's orbit (black line). We use EUHFORIA data to compute the HCS location at 1 AU by separating positive and negative polarities. We, unfortunately, do not have a way to measure the HCS position around Earth (it is possible to determine its shape around 4 solar radii using SOHO/LASCO data, see \cite{Poirier2021}, but not further without using theoretical assumptions). However, we can check whether the crossings between Earth orbit and the HCS correspond to polarity switches. Our HCS positioning is off during this period if they do not match. To better illustrate this potential mismatch, the middle column is essentially the same plot, except that this time, the polarity at Earth positions comes from the MHD simulations. It can then be directly compared to the measured in situ polarity to see which dates are off. Finally, the right column quantifies the agreement between in situ and simulated polarity for each date (in green when there is agreement, in red otherwise). From this final plot, we can compute a percentage of agreement between data and simulations, quantifying the quality of our global polarity estimate and, thus, our chances of getting the right magnetic connectivity for an event during this period. 

Using the latter, we can determine that we have a 69\% agreement for event 1, a 36\% agreement for event 2, a 68\% agreement for event 3 and a 69\% agreement for event 4. This shows the consistency of our MHD simulations and that, for most of these cases, we are close to being able to guarantee the right connectivity at 70\%. Interestingly, even though event 2 has the least good estimation of the HCS, we still recover good magnetic connectivity. In contrast, events 1 and 3 have good polarity estimation despite lower scores for CH connectivity. We can then look at the left column of Figure~\ref{fig:exp_samen} to understand these results further. 

For event 1 (top line), the explanation is relatively straightforward: we can see on the left panel that the negative polarity (in blue) was measured at Earth sooner than in our simulations. This slight delay of 24 hours explains why we have residual connectivity to the northern CH. Then, it is impossible to distinguish between the contributions of the equatorial and southern CH, which have the same polarity. The fact that the equatorial CH is smaller than observed is still probably the biggest issue. 

For event 2 (second line), we can see that the polarity around the event in the simulations is positive (in red) as it should be, which explains why we get the event right. However, right after the event, we switch early to a negative polarity (in blue) due to the crossing of the HCS. This switch is close to what is observed in the data, where this patchy CH leads to rapid switches between positive and negative polarities. In this case, the uncertainty of the simulation is accentuated by the fact that we are close to the end of the Carrington rotation, which can cause discrepancies in the simulation results. Hence, this confirms that it was a very difficult case but that we could still recover the global trends. 

For event 3 (third line), we can see that the HCS positioning for the blue negative polarity between 31/07 and 04/08 is off in our simulation, with the blue positive polarity starting too early on 01/08. This should help us recover the right CH. However, we still get some magnetic field lines from the negative southern CH. This means that because of the large extended northern CH, field lines are more twisted and thus, some isolated positive field lines may be close to the Earth location in our simulation. This means that, in this case, we do recover the right magnetic connectivity at the exact Earth location. Still, the surrounding uncertainty is high, which cannot be tested similarly for the observations.

For event 4 (bottom line), the HSS event happens during a long period of negative polarity, which is well reproduced in our simulation. Because of the larger latitudinal amplitude of the HCS (between -30 and 60 degrees here, compared to -10 and 15 degrees in previous cases), it may be easier to recover the right polarity for MHD simulations. This could mean that the maximum number of activity cases may be more reliable than the minimum number of activity cases because the crossing of the HCS is better defined and happens less often for the Earth's location. This conclusion is consistent with the findings by \citet{Badman2023} for Parker Solar Probe data. This explains why we recover the right polarity. We cannot then distinguish between the various negative CHs.

In conclusion, with this additional analysis, we can quantify that, on average, our MHD simulations provide the proper magnetic connectivity with an almost 70\% rate. We can explain uncertainties in our results because of errors in the HCS positioning (event 1). We can also proceed to a more in-depth analysis and show that the polarity estimate does not guarantee perfect magnetic connectivity: event 2 is a very difficult case and still provides a good result, while event 3 should have been easy and generated mixed results. This analysis shows that the maximum of activity cases may be easier to compute for MHD simulations due to the more extended shape of the HCS. Cases at minimum activity indeed show more polarity switches, requiring higher resolution and maybe even more physical (non-ideal MHD for the reconnection) and numerical (time-dependent simulations) features to be perfectly captured. 

\section{Conclusions}
\label{sec:Conclusion}

This study follows the recent developments surrounding the determination of the Sun-to-Earth magnetic connectivity, which is crucial for understanding solar wind events and anticipating space weather events. Traditionally, the solar magnetic connectivity is determined using a two-step ballistic mapping, combining a PFSS extrapolation close to the Sun with a Parker spiral approximation close to Earth. However, this method uses a lot of free parameters that can significantly influence the final result. Here, we want to use another approach based on MHD simulations, which are more self-consistent and physics-based and can now perform on an operational scale. To do so, we combine the coronal model COCONUT with the heliospheric model EUHFORIA and trace the magnetic field lines from the Earth to the Sun. We can define spatial uncertainty by probing locations close to Earth and temporal uncertainty by computing the magnetic connectivity averaged over a few days before the event. Using the model's default parameters, we check whether we can recover the correct magnetic connectivity for various standard events. Four cases were chosen, three during a solar minimum and one during a solar maximum, already investigated in \citet{Reiss2021} and \citet{koukras2022}. These cases correspond to HSS events, which is the most reliable magnetic connectivity validation we can get since there are no direct observations of the magnetic connectivity. 

For all four selected events, we always find at least a partial overlap with the assumed CH of origin (19\% for event 1, 100\% for event 2, 45\% for event 3 and 100\% for event 4). This is better than the two-step ballistic method, where event 2 leads to a completely wrong connection. The fact that we also include other CHs comes from the spatial and temporal uncertainty. We looked at the global polarity of the simulation for Earth over the full Carrington rotation to understand these results better. We found that, on average, MHD simulations provide a very good polarity estimation (69\% agreement with real data for event 1, 36\% for event 2, 68\% for event 3 and 69\% for event 4). For event 1, it is the temporal uncertainty that yields mixed results due to a delay of 24 hours for the HCS crossing, while for event 3 it is the spatial uncertainty that yields mixed results due to a complex open magnetic structure. Challenging cases still provide good results: event 2, even though it is a complex case with a patchy CH and challenging magnetic connectivity; event 4, even though it is a maximum activity case. This helps us find the benefits and limits of this method of computation for the connectivity: minimum activity cases appear to be more challenging because of multiple recurrent crossings of the HCS, which are difficult to reproduce accurately for ideal MHD simulations; maximum activity cases appear easier for MHD models because of the latitudinal extent of the HCS. Of course, this analysis would require more statistics to confirm these trends, so we need more confirmed connectivity events for validation. 

In conclusion, we have demonstrated that we can use MHD models to compute the magnetic connectivity between the Sun and the Earth. The connectivity quality will rely on the good match between the simulation and observations (which calls for more automated validation techniques for both coronal and heliospheric simulations). Still, even with default parameters for four different dates between 2012 and 2020, we have provided at least a partial overlap with the correct CH for all cases. It is just as performant as the thoroughly fine-tuned two-step ballistic method. Still, it yields more information about the global magnetic field polarity configuration (which can help provide feedback on improving the simulation), does not require any adjustment, and offers significant room for improvement by developing better models. A more self-consistent inclusion of the solar rotation for example could be added to reduce the spatial and temporal uncertainty of our estimations. The COCONUT model has already been enhanced with the inclusion of multiple fluids \citep{Brchnelova2023_mf}, while the EUHFORIA model has been upgraded to AMR features with its new version ICARUS \citep{Verbeke2022, Baratashvili2024}. In particular, MHD models would need to use a better spatial resolution to get more small-scale changes in the HCS and include non-ideal MHD effects to have a more physics-based reconnection in the HCS. Ideally, models should even be time-dependent to remove the need for a temporal uncertainty \citep{Lionello2023}. Studies with other MHD models could be carried out to confirm these trends. The next step of this study will focus on sources of uncertainty for connectivity other than the models themselves; in particular, we will look at the impact of the input magnetic map and quantify the effect of our final estimates. 


\begin{acknowledgements}
The authors thank Rui Pinto, Predictive Science Inc. and the ISWAT S2-05 team for valuable discussions. 
SK and SP acknowledge support from the projects C14/19/089 (C1 project Internal Funds KU Leuven), G0B5823N and G002523N (WEAVE) (FWO-Vlaanderen), 4000145223 (SIDC Data Exploitation (SIDEX2), ESA Prodex), and Belspo project B2/191/P1/SWiM.   
We used the VSC – Flemish Supercomputer Center infrastructure for the computations, funded by the Hercules Foundation and the Flemish Government – department EWI.
HMI data are courtesy of the Joint Science Operations Center (JSOC) Science Data Processing team at Stanford University.
We thank the ACE SWEPAM instrument team and the ACE Science Center for providing the ACE data.
\end{acknowledgements}

\bibliographystyle{aa}
\bibliography{references}

\begin{thebibliography}{73}
\expandafter\ifx\csname natexlab\endcsname\relax\def\natexlab#1{#1}\fi

\bibitem[{{Altschuler} \& {Newkirk}(1969)}]{Altschuler1969}
{Altschuler}, M.~D. \& {Newkirk}, G. 1969, \solphys, 9, 131

\bibitem[{Annex {et~al.}(2020)Annex, Pearson, Seignovert, Carcich, Eichhorn,
  Mapel, von Forstner, McAuliffe, del Rio, Berry, Aye, Stefko, de~Val-Borro,
  Kulumani, \& ya~Murakami}]{andrew2020}
Annex, A.~M., Pearson, B., Seignovert, B., {et~al.} 2020, Journal of Open
  Source Software, 5, 2050

\bibitem[{{Badman} {et~al.}(2020){Badman}, {Bale}, {Mart{\'\i}nez Oliveros},
  {Panasenco}, {Velli}, {Stansby}, {Buitrago-Casas}, {R{\'e}ville}, {Bonnell},
  {Case}, {Dudok de Wit}, {Goetz}, {Harvey}, {Kasper}, {Korreck}, {Larson},
  {Livi}, {MacDowall}, {Malaspina}, {Pulupa}, {Stevens}, \&
  {Whittlesey}}]{Badman2020}
{Badman}, S.~T., {Bale}, S.~D., {Mart{\'\i}nez Oliveros}, J.~C., {et~al.} 2020,
  \apjs, 246, 23

\bibitem[{Badman {et~al.}(2022)Badman, Brooks, Poirier, Warren, Petrie,
  Rouillard, Arge, Bale, de~Pablos~Agüero, Harra, Jones, Kouloumvakos, Riley,
  Panasenco, Velli, \& Wallace}]{Badman2022}
Badman, S.~T., Brooks, D.~H., Poirier, N., {et~al.} 2022, The Astrophysical
  Journal, 932, 135

\bibitem[{{Badman} {et~al.}(2023){Badman}, {Riley}, {Jones}, {Kim}, {Allen},
  {Arge}, {Bale}, {Henney}, {Kasper}, {Mostafavi}, {Pogorelov}, {Raouafi},
  {Stevens}, \& {Verniero}}]{Badman2023}
{Badman}, S.~T., {Riley}, P., {Jones}, S.~I., {et~al.} 2023, Journal of
  Geophysical Research (Space Physics), 128, e2023JA031359

\bibitem[{{Baker} {et~al.}(2023){Baker}, {D{\'e}moulin}, {Yardley},
  {Mihailescu}, {van Driel-Gesztelyi}, {D'Amicis}, {Long}, {To}, {Owen},
  {Horbury}, {Brooks}, {Perrone}, {French}, {James}, {Janvier}, {Matthews},
  {Stangalini}, {Valori}, {Smith}, {Cuadrado}, {Peter}, {Schuehle}, {Harra},
  {Barczynski}, {Berghmans}, {Zhukov}, {Rodriguez}, \& {Verbeeck}}]{Baker2023}
{Baker}, D., {D{\'e}moulin}, P., {Yardley}, S.~L., {et~al.} 2023, \apj, 950, 65

\bibitem[{{Baratashvili} \& {Poedts}(2024)}]{Baratashvili2024}
{Baratashvili}, T. \& {Poedts}, S. 2024, \aap, 683, A81

\bibitem[{Beck(2010)}]{John2010}
Beck, J. 2010, HMI \& WCS COORDINATES, PROJECTIONS AND ARRAYS FOR DUMMIES

\bibitem[{{Brchnelova} {et~al.}(2022{\natexlab{a}}){Brchnelova}, {Ku{\'z}ma},
  {Perri}, {Lani}, \& {Poedts}}]{Brchnelova2022b}
{Brchnelova}, M., {Ku{\'z}ma}, B., {Perri}, B., {Lani}, A., \& {Poedts}, S.
  2022{\natexlab{a}}, \apjs, 263, 18

\bibitem[{{Brchnelova} {et~al.}(2023){Brchnelova}, {Ku{\'z}ma}, {Zhang},
  {Lani}, \& {Poedts}}]{Brchnelova2023_mf}
{Brchnelova}, M., {Ku{\'z}ma}, B., {Zhang}, F., {Lani}, A., \& {Poedts}, S.
  2023, \aap, 678, A117

\bibitem[{{Brchnelova} {et~al.}(2022{\natexlab{b}}){Brchnelova}, {Zhang},
  {Leitner}, {Perri}, {Lani}, \& {Poedts}}]{Brchnelova2022}
{Brchnelova}, M., {Zhang}, F., {Leitner}, P., {et~al.} 2022{\natexlab{b}},
  Journal of Plasma Physics, 88, 905880205

\bibitem[{{Caplan} {et~al.}(2016){Caplan}, {Downs}, \& {Linker}}]{Caplan2016}
{Caplan}, R.~M., {Downs}, C., \& {Linker}, J.~A. 2016, \apj, 823, 53

\bibitem[{{Chorin}(1997)}]{chorin1997}
{Chorin}, A.~J. 1997, Journal of Computational Physics, 135, 118

\bibitem[{{Cid} {et~al.}(2012){Cid}, {Cremades}, {Aran}, {Mandrini},
  {Sanahuja}, {Schmieder}, {Menvielle}, {Rodriguez}, {Saiz}, {Cerrato},
  {Dasso}, {Jacobs}, {Lathuillere}, \& {Zhukov}}]{Cid2012}
{Cid}, C., {Cremades}, H., {Aran}, A., {et~al.} 2012, Journal of Geophysical
  Research (Space Physics), 117, A11102

\bibitem[{{Cranmer} {et~al.}(2017){Cranmer}, {Gibson}, \&
  {Riley}}]{Cranmer2017}
{Cranmer}, S.~R., {Gibson}, S.~E., \& {Riley}, P. 2017, \ssr, 212, 1345

\bibitem[{{da Silva} {et~al.}(2023){da Silva}, {Wallace}, {Arge}, \&
  {Jones}}]{DaSilva2023}
{da Silva}, D.~E., {Wallace}, S., {Arge}, C.~N., \& {Jones}, S. 2023, Space
  Weather, 21, e2023SW003554

\bibitem[{{Dakeyo} {et~al.}(2024){Dakeyo}, {Badman}, {Rouillard},
  {R{\'e}ville}, {Verscharen}, {D{\'e}moulin}, \& {Maksimovic}}]{Dakeyo2024}
{Dakeyo}, J.~B., {Badman}, S.~T., {Rouillard}, A.~P., {et~al.} 2024, \aap, 686,
  A12

\bibitem[{Dedner {et~al.}(2002)Dedner, Kemm, Kröner, Munz, Schnitzer, \&
  Wesenberg}]{Dedner2002}
Dedner, A., Kemm, F., Kröner, D., {et~al.} 2002, Journal of Computational
  Physics, 175, 645

\bibitem[{{Grandin} {et~al.}(2019){Grandin}, {Aikio}, \&
  {Kozlovsky}}]{Grandin2019}
{Grandin}, M., {Aikio}, A.~T., \& {Kozlovsky}, A. 2019, Journal of Geophysical
  Research (Space Physics), 124, 3871

\bibitem[{{Hathaway}(2015)}]{Hathaway2015}
{Hathaway}, D.~H. 2015, Living Reviews in Solar Physics, 12, 4

\bibitem[{{Heber} \& {Potgieter}(2006)}]{Heber2006}
{Heber}, B. \& {Potgieter}, M.~S. 2006, \ssr, 127, 117

\bibitem[{{Helioviewer Project}(2023)}]{helioviewer}
{Helioviewer Project}. 2023, Helioviewer: Solar and Heliospheric Image Browser,
  accessed: 2024-08-13

\bibitem[{{Hoeksema} {et~al.}(1983){Hoeksema}, {Wilcox}, \&
  {Scherrer}}]{Hoeksema1983}
{Hoeksema}, J.~T., {Wilcox}, J.~M., \& {Scherrer}, P.~H. 1983, \jgr, 88, 9910

\bibitem[{{Kissmann} \& {Pomoell}(2012)}]{Kissmann2012}
{Kissmann}, R. \& {Pomoell}, J. 2012, SIAM Journal on Scientific Computing, 34,
  A763

\bibitem[{{Kohl} {et~al.}(2006){Kohl}, {Noci}, {Cranmer}, \&
  {Raymond}}]{Kohl2006}
{Kohl}, J.~L., {Noci}, G., {Cranmer}, S.~R., \& {Raymond}, J.~C. 2006, \aapr,
  13, 31

\bibitem[{{Koukras} {et~al.}(2022){Koukras}, {Dolla}, \&
  {Keppens}}]{koukras2022}
{Koukras}, A., {Dolla}, L., \& {Keppens}, R. 2022, in SHINE 2022 Workshop, 68

\bibitem[{{Krieger} {et~al.}(1973){Krieger}, {Timothy}, \&
  {Roelof}}]{Krieger1973}
{Krieger}, A.~S., {Timothy}, A.~F., \& {Roelof}, E.~C. 1973, \solphys, 29, 505

\bibitem[{{Ku{\'z}ma} {et~al.}(2023){Ku{\'z}ma}, {Brchnelova}, {Perri},
  {Baratashvili}, {Zhang}, {Lani}, \& {Poedts}}]{kuzma2023}
{Ku{\'z}ma}, B., {Brchnelova}, M., {Perri}, B., {et~al.} 2023, \apj, 942, 31

\bibitem[{{Landi} {et~al.}(2012){Landi}, {Gruesbeck}, {Lepri}, \&
  {Zurbuchen}}]{Landi2012}
{Landi}, E., {Gruesbeck}, J.~R., {Lepri}, S.~T., \& {Zurbuchen}, T.~H. 2012,
  \apj, 750, 159

\bibitem[{Lani {et~al.}(2013)Lani, Villedieu, Bensassi, Koloszar, Panesi, \&
  Yalim}]{Lani2013}
Lani, A., Villedieu, N., Bensassi, K., {et~al.} 2013

\bibitem[{{Lanzerotti}(2001)}]{Lanzerotti2001}
{Lanzerotti}, L.~J. 2001, Geophysical Monograph Series, 125, 11

\bibitem[{{Lavraud} \& {Rouillard}(2014)}]{Lavraud2014}
{Lavraud}, B. \& {Rouillard}, A. 2014, in Nature of Prominences and their Role
  in Space Weather, ed. B.~{Schmieder}, J.-M. {Malherbe}, \& S.~T. {Wu}, Vol.
  300, 273--284

\bibitem[{{Lee} {et~al.}(2011){Lee}, {Luhmann}, {Hoeksema}, {Sun}, {Arge}, \&
  {de Pater}}]{Lee2011}
{Lee}, C.~O., {Luhmann}, J.~G., {Hoeksema}, J.~T., {et~al.} 2011, \solphys,
  269, 367

\bibitem[{{Linker} \& {Miki{\'c}}(1997)}]{Linker1997}
{Linker}, J.~A. \& {Miki{\'c}}, Z. 1997, Geophysical Monograph Series, 99, 269

\bibitem[{{Lionello} {et~al.}(2023){Lionello}, {Downs}, {Mason}, {Linker},
  {Caplan}, {Riley}, {Titov}, \& {DeRosa}}]{Lionello2023}
{Lionello}, R., {Downs}, C., {Mason}, E.~I., {et~al.} 2023, \apj, 959, 77

\bibitem[{{Macneil} {et~al.}(2022){Macneil}, {Owens}, {Finley}, \&
  {Matt}}]{MacNeil2022}
{Macneil}, A.~R., {Owens}, M.~J., {Finley}, A.~J., \& {Matt}, S.~P. 2022,
  \mnras, 509, 2390

\bibitem[{{McComas} {et~al.}(2000){McComas}, {Barraclough}, {Funsten},
  {Gosling}, {Santiago-Mu{\~n}oz}, {Skoug}, {Goldstein}, {Neugebauer}, {Riley},
  \& {Balogh}}]{McComas2000}
{McComas}, D.~J., {Barraclough}, B.~L., {Funsten}, H.~O., {et~al.} 2000, \jgr,
  105, 10419

\bibitem[{{Moffatt}(1978)}]{Moffatt1978}
{Moffatt}, H.~K. 1978, {Magnetic field generation in electrically conducting
  fluids}

\bibitem[{{Neugebauer} {et~al.}(1998){Neugebauer}, {Forsyth}, {Galvin},
  {Harvey}, {Hoeksema}, {Lazarus}, {Lepping}, {Linker}, {Mikic}, {Steinberg},
  {von Steiger}, {Wang}, \& {Wimmer-Schweingruber}}]{Neugebauer1998}
{Neugebauer}, M., {Forsyth}, R.~J., {Galvin}, A.~B., {et~al.} 1998, \jgr, 103,
  14587

\bibitem[{{Nolte} \& {Roelof}(1973)}]{Nolte1973}
{Nolte}, J.~T. \& {Roelof}, E.~C. 1973, \solphys, 33, 241

\bibitem[{{Odstrcil} {et~al.}(2004){Odstrcil}, {Riley}, \&
  {Zhao}}]{Odstrcil2004}
{Odstrcil}, D., {Riley}, P., \& {Zhao}, X.~P. 2004, Journal of Geophysical
  Research (Space Physics), 109, A02116

\bibitem[{Owens \& Forsyth(2013)}]{Owens2013}
Owens, M.~J. \& Forsyth, R.~J. 2013, Living Reviews in Solar Physics, 10, 5

\bibitem[{{Parker}(1958)}]{Parker1958}
{Parker}, E.~N. 1958, \apj, 128, 664

\bibitem[{{Parker}(1993)}]{Parker1993}
{Parker}, E.~N. 1993, \apj, 408, 707

\bibitem[{{Peleikis} {et~al.}(2017){Peleikis}, {Kruse}, {Berger}, \&
  {Wimmer-Schweingruber}}]{Peleikis2017}
{Peleikis}, T., {Kruse}, M., {Berger}, L., \& {Wimmer-Schweingruber}, R. 2017,
  \aap, 602, A24

\bibitem[{Perri {et~al.}(2023)Perri, Kuźma, Brchnelova, Baratashvili, Zhang,
  Leitner, Lani, \& Poedts}]{Perri2023}
Perri, B., Kuźma, B., Brchnelova, M., {et~al.} 2023, The Astrophysical
  Journal, 943, 124

\bibitem[{Perri {et~al.}(2022)Perri, Leitner, Brchnelova, Baratashvili,
  Ku{\'{z}}ma, Zhang, Lani, \& Poedts}]{Perri2022}
Perri, B., Leitner, P., Brchnelova, M., {et~al.} 2022, The Astrophysical
  Journal, 936, 19

\bibitem[{{Poedts} {et~al.}(2020){Poedts}, {Kochanov}, {Lani}, {Scolini},
  {Verbeke}, {Hosteaux}, {Chan{\'e}}, {Deconinck}, {Mihalache}, {Diet},
  {Heynderickx}, {De Keyser}, {De Donder}, {Crosby}, {Echim}, {Rodriguez},
  {Vansintjan}, {Verstringe}, {Mampaey}, {Horne}, {Glauert}, {Jiggens}, {Keil},
  {Glover}, {Deprez}, \& {Luntama}}]{Poedts2020}
{Poedts}, S., {Kochanov}, A., {Lani}, A., {et~al.} 2020, Journal of Space
  Weather and Space Climate, 10, 14

\bibitem[{{Poirier} {et~al.}(2021){Poirier}, {Rouillard}, {Kouloumvakos},
  {Przybylak}, {Fargette}, {Pobeda}, {R{\'e}ville}, {Pinto}, {Indurain}, \&
  {Alexandre}}]{Poirier2021}
{Poirier}, N., {Rouillard}, A.~P., {Kouloumvakos}, A., {et~al.} 2021, Frontiers
  in Astronomy and Space Sciences, 8, 84

\bibitem[{Pomoell \& Poedts(2018)}]{Pomoell2018}
Pomoell, J. \& Poedts, S. 2018, J. Space Weather Space Clim., 8, A35

\bibitem[{{Pomoell} \& {Vainio}(2012)}]{Pomoell2012}
{Pomoell}, J. \& {Vainio}, R. 2012, \apj, 745, 151

\bibitem[{{Pulkkinen}(2007)}]{Pulkkinen2007}
{Pulkkinen}, T. 2007, Living Reviews in Solar Physics, 4, 1

\bibitem[{Reiss {et~al.}(2021)Reiss, Muglach, Möstl, Arge, Bailey, Delouille,
  Garton, Hamada, Hofmeister, Illarionov, Jarolim, Kirk, Kosovichev, Krista,
  Lee, Lowder, MacNeice, Veronig, \& Team}]{Reiss2021}
Reiss, M.~A., Muglach, K., Möstl, C., {et~al.} 2021, The Astrophysical
  Journal, 913, 28

\bibitem[{{Riley} {et~al.}(2006){Riley}, {Linker}, {Miki{\'c}}, {Lionello},
  {Ledvina}, \& {Luhmann}}]{Riley2006}
{Riley}, P., {Linker}, J.~A., {Miki{\'c}}, Z., {et~al.} 2006, \apj, 653, 1510

\bibitem[{{Riley} {et~al.}(2018){Riley}, {Mays}, {Andries}, {Amerstorfer},
  {Biesecker}, {Delouille}, {Dumbovi{\'c}}, {Feng}, {Henley}, {Linker},
  {M{\"o}stl}, {Nu{\~n}ez}, {Pizzo}, {Temmer}, {Tobiska}, {Verbeke}, {West}, \&
  {Zhao}}]{Riley2018}
{Riley}, P., {Mays}, M.~L., {Andries}, J., {et~al.} 2018, Space Weather, 16,
  1245

\bibitem[{{Roelof} \& {Krimigis}(1973)}]{Roelof1973}
{Roelof}, E.~C. \& {Krimigis}, S.~M. 1973, \jgr, 78, 5375

\bibitem[{{Rouillard} {et~al.}(2020){Rouillard}, {Pinto}, {Vourlidas}, {De
  Groof}, {Thompson}, {Bemporad}, {Dolei}, {Indurain}, {Buchlin}, {Sasso},
  {Spadaro}, {Dalmasse}, {Hirzberger}, {Zouganelis}, {Strugarek}, {Brun},
  {Alexandre}, {Berghmans}, {Raouafi}, {Wiegelmann}, {Pagano}, {Arge},
  {Nieves-Chinchilla}, {Lavarra}, {Poirier}, {Amari}, {Aran}, {Andretta},
  {Antonucci}, {Anastasiadis}, {Auch{\`e}re}, {Bellot Rubio}, {Nicula},
  {Bonnin}, {Bouchemit}, {Budnik}, {Caminade}, {Cecconi}, {Carlyle}, {Cernuda},
  {Davila}, {Etesi}, {Espinosa Lara}, {Fedorov}, {Fineschi}, {Fludra},
  {G{\'e}not}, {Georgoulis}, {Gilbert}, {Giunta}, {Gomez-Herrero}, {Guest},
  {Haberreiter}, {Hassler}, {Henney}, {Howard}, {Horbury}, {Janvier}, {Jones},
  {Kozarev}, {Kraaikamp}, {Kouloumvakos}, {Krucker}, {Lagg}, {Linker},
  {Lavraud}, {Louarn}, {Maksimovic}, {Maloney}, {Mann}, {Masson}, {M{\"u}ller},
  {{\"O}nel}, {Osuna}, {Orozco Suarez}, {Owen}, {Papaioannou},
  {P{\'e}rez-Su{\'a}rez}, {Rodriguez-Pacheco}, {Parenti}, {Pariat}, {Peter},
  {Plunkett}, {Pomoell}, {Raines}, {Riethm{\"u}ller}, {Rich}, {Rodriguez},
  {Romoli}, {Sanchez}, {Solanki}, {St Cyr}, {Straus}, {Susino}, {Teriaca}, {del
  Toro Iniesta}, {Ventura}, {Verbeeck}, {Vilmer}, {Warmuth}, {Walsh}, {Watson},
  {Williams}, {Wu}, \& {Zhukov}}]{Rouillard2020}
{Rouillard}, A.~P., {Pinto}, R.~F., {Vourlidas}, A., {et~al.} 2020, \aap, 642,
  A2

\bibitem[{{Schatten}(1971)}]{Schatten1971}
{Schatten}, K.~H. 1971, Cosmic Electrodynamics, 2, 232

\bibitem[{{Schatten} {et~al.}(1969){Schatten}, {Wilcox}, \&
  {Ness}}]{Schatten1969}
{Schatten}, K.~H., {Wilcox}, J.~M., \& {Ness}, N.~F. 1969, \solphys, 6, 442

\bibitem[{{Schrijver} \& {De Rosa}(2003)}]{schrijver2003}
{Schrijver}, C.~J. \& {De Rosa}, M.~L. 2003, \solphys, 212, 165

\bibitem[{{Schrijver} {et~al.}(2015){Schrijver}, {Kauristie}, {Aylward},
  {Denardini}, {Gibson}, {Glover}, {Gopalswamy}, {Grande}, {Hapgood},
  {Heynderickx}, {Jakowski}, {Kalegaev}, {Lapenta}, {Linker}, {Liu},
  {Mandrini}, {Mann}, {Nagatsuma}, {Nandy}, {Obara}, {Paul O'Brien}, {Onsager},
  {Opgenoorth}, {Terkildsen}, {Valladares}, \& {Vilmer}}]{Schrijver2015_cospar}
{Schrijver}, C.~J., {Kauristie}, K., {Aylward}, A.~D., {et~al.} 2015, Advances
  in Space Research, 55, 2745

\bibitem[{{Singh} {et~al.}(2022){Singh}, {Kim}, {Pogorelov}, \&
  {Arge}}]{Singh2022}
{Singh}, T., {Kim}, T.~K., {Pogorelov}, N.~V., \& {Arge}, C.~N. 2022, \apj,
  933, 123

\bibitem[{{Snyder} {et~al.}(1963){Snyder}, {Neugebauer}, \& {Rao}}]{Snyder1963}
{Snyder}, C.~W., {Neugebauer}, M., \& {Rao}, U.~R. 1963, \jgr, 68, 6361

\bibitem[{Stansby {et~al.}(2020)Stansby, Yeates, \& Badman}]{Stansby2020}
Stansby, D., Yeates, A., \& Badman, S.~T. 2020, Journal of Open Source
  Software, 5, 2732

\bibitem[{{Sullivan} \& {Kaszynski}(2019)}]{Sullivan2019}
{Sullivan}, C. \& {Kaszynski}, A. 2019, The Journal of Open Source Software, 4,
  1450

\bibitem[{{Temmer}(2021)}]{Temmer2021}
{Temmer}, M. 2021, Living Reviews in Solar Physics, 18, 4

\bibitem[{{Thompson}(2006)}]{Thompson2006}
{Thompson}, W.~T. 2006, \aap, 449, 791

\bibitem[{{Verbanac} {et~al.}(2011){Verbanac}, {Vr{\v{s}}nak}, {Veronig}, \&
  {Temmer}}]{Verbanac2011}
{Verbanac}, G., {Vr{\v{s}}nak}, B., {Veronig}, A., \& {Temmer}, M. 2011, \aap,
  526, A20

\bibitem[{{Verbeke, C.} {et~al.}(2022){Verbeke, C.}, {Baratashvili, T.}, \&
  {Poedts, S.}}]{Verbeke2022}
{Verbeke, C.}, {Baratashvili, T.}, \& {Poedts, S.} 2022, A\&A, 662, A50

\bibitem[{{Vr{\v{s}}nak} {et~al.}(2007){Vr{\v{s}}nak}, {Temmer}, \&
  {Veronig}}]{Vrsnak2007}
{Vr{\v{s}}nak}, B., {Temmer}, M., \& {Veronig}, A.~M. 2007, \solphys, 240, 331

\bibitem[{{Yardley} {et~al.}(2024){Yardley}, {Brooks}, {D'Amicis}, {Owen},
  {Long}, {Baker}, {D{\'e}moulin}, {Owens}, {Lockwood}, {Mihailescu}, {Coburn},
  {Dewey}, {M{\"u}ller}, {Suen}, {Ngampoopun}, {Louarn}, {Livi}, {Lepri},
  {Fludra}, {Haberreiter}, \& {Sch{\"u}hle}}]{Yardley2024}
{Yardley}, S.~L., {Brooks}, D.~H., {D'Amicis}, R., {et~al.} 2024, Nature
  Astronomy

\bibitem[{{Zhang} {et~al.}(2021){Zhang}, {Temmer}, {Gopalswamy}, {Malandraki},
  {Nitta}, {Patsourakos}, {Shen}, {Vr{\v{s}}nak}, {Wang}, {Webb}, {Desai},
  {Dissauer}, {Dresing}, {Dumbovi{\'c}}, {Feng}, {Heinemann}, {Laurenza},
  {Lugaz}, \& {Zhuang}}]{Zhang2021}
{Zhang}, J., {Temmer}, M., {Gopalswamy}, N., {et~al.} 2021, Progress in Earth
  and Planetary Science, 8, 56

\bibitem[{{Zirker}(1977)}]{Zirker1977}
{Zirker}, J.~B. 1977, Reviews of Geophysics and Space Physics, 15, 257

\end{thebibliography}

\clearpage
\onecolumn

\end{document}